\documentclass[%
 reprint,
 amsmath,amssymb,
 aps,
]{revtex4-2}

\usepackage{graphicx}
\usepackage{dcolumn}
\usepackage{bm}
\usepackage{appendix}

\usepackage[colorlinks=true, linkcolor=blue, citecolor=blue]{hyperref}

\newcommand\authormark[1]{\textsuperscript{#1}}
\newcommand{\manuscripttitle}{Resonance-enhanced integrated acousto-optic beam steering}
\newcommand{\authorlist}{%
  Yue Yu\authormark{1}, Qixuan Lin\authormark{1}, Shucheng Fang\authormark{1}, Joseph G. Thomas\authormark{2}, Yibing Zhou\authormark{1}, Zichen Xi\authormark{2}, Jun Ji\authormark{2}, Yizheng Zhu\authormark{2}, Linbo Shao\authormark{2,3}, Bingzhao Li\authormark{1}, and Mo Li\authormark{1,4}%
}

\newcommand{\affiliations}{%
  \authormark{1} Department of Electrical and Computer Engineering, University of Washington, Seattle, WA, USA\\
  \authormark{2}Bradley Department of Electrical and Computer Engineering, Virginia Tech, Blacksburg, VA, USA\\
  \authormark{3}Center for Quantum Information Science and Engineering (VTQ), Virginia Tech, Blacksburg, VA, USA\\
  \authormark{4}Department of Physics, University of Washington, Seattle, WA, USA%
}
\newcommand{\correspondingauthor}{moli96@uw.edu}

\begin{document}

\title{\manuscripttitle}

\author{\authorlist}
\email{\correspondingauthor}

\affiliation{\affiliations}

\begin{abstract}
Optical beam steering is a key technology for free-space optical communication, sensing, and imaging. Mechanical beam steering systems suffer from limited scanning speed and bulky form factors, while existing solid-state solutions rely on pixelated synthetic aperture that requires complex fabrication and control architectures. Integrated acousto-optic beam steering (AOBS) is an emerging technology that enables continuous one-dimensional beam steering using integrated acoustic transducers and fixed-wavelength laser sources. Here, we integrate AOBS with an optical ring resonator on the same thin-film lithium niobate (TFLN) platform to significantly enhance beam steering efficiency and system functionality. The resulting device achieves a resonance-enhanced beam steering efficiency of up to $26\%$ and a field of view of $18^\circ$. Moreover, by leveraging integrated electro-optic control, we dynamically lock the ring-resonator's resonance to a chirped laser frequency, enabling frequency-modulated continuous-wave (FMCW) LiDAR operation. By combining lithium niobate's piezoelectric and electro-optic properties, this work establishes a compact, efficient, and scalable beam-steering platform with co-integrated acousto-optic modulation and electro-optic control for multifunctional applications.
\end{abstract}

\maketitle

\section{Introduction}
Optical beam steering is a key technology is a key enabling technology for a wide range of optical applications, including light detection and ranging (LiDAR)~\cite{wehr1999airborne,schwarz2010mapping}, free-space optical communication~\cite{khalighi2014survey, saghaye2019imaging}, bioimaging~\cite{salome2006ultrafast, von1983beam}, optical tweezers~\cite{dufresne1998optical, curtis2002dynamic}, and optical control of quantum systems~\cite{wang2015coherent, naegerl1999laser, endres2016atom}. Conventional non-mechanical beam steering approaches, such as optical phased arrays (OPA)~\cite{sun2013large, chul2020chip, liu2022silicon}, spatial light modulators (SLM)~\cite{zhang2024scaled}, focal-plane switch arrays (FPSA)~\cite{zhang2022large, rogers2021universal}, and digital micromirror devices (DMD)~\cite{zupancic2016ultra, shih2021reprogrammable}, rely on synthetic apertures composed of a large number of wavelength-scale pixels. Achieving wide steering angles and high angular resolution with these platforms requires increasingly larger arrays of smaller pixels, thereby demanding more sophisticated peripheral control circuitry, which poses significant challenges in terms of scalability, power consumption, and system complexity. These challenges become even more severe when the application requires steering multiple beams simultaneously, such as high-frame-rate LiDAR, multi-channel free-space optical communication, and scalable quantum computing using atomic and ionic qubits. An alternative approach to beam steering exploits dispersive optical elements to diffract different optical frequencies to different directions~\cite{riemensberger2020massively,trocha2018ultrafast}. While this principle enables beam steering using wavelength tuning, it typically requires broadband or rapidly tunable laser sources, which are often complex and costly, and incompatible with many applications.

Acousto-optic beam steering (AOBS) is an emerging technology with the potential to address the aforementioned challenges due to its distinct operating principle. AOBS diffracts light using acoustic waves in the material, which generates a dynamically tunable index grating  ~\cite{brillouin1922diffusion, debye1932scattering}. The spatial period of the grating and the diffraction angle in AOBS are controlled by the acoustic frequency, while the scattering strength is controlled by the acoustic amplitude, allowing beam steering with amplitude control. Conventional bulk acousto-optic devices have long been used as deflectors, modulators, and frequency shifters~\cite{gordon1966review}, but their performances are limited by the relatively low acoustic frequencies and large device sizes.

Recent advances in integrated photonics and optomechanics have enabled guided-wave acousto-optic devices that confine both optical and acoustic waves within planar~\cite{shao2020integrated} and waveguide~\cite{zhang2024integrated, chen2025intermodal} structures. The strong confinement substantially increases the interaction between acoustic and optical waves in a compact device footprint. Compared with other solid-state beam steering techniques that use pixelated optical apertures, integrated AOBS~\cite{li2023frequency, lin2025optical} offer several unique advantages: they feature a continuous optical aperture, require only one acoustic transducer for each channel, and allow simultaneous multi-beam steering by driving each channel with multiple acoustic tones. Furthermore, the amplitude and phase of each steered beam can be independently controlled with high bandwidth, making AOBS a highly versatile and scalable technology for many applications.

\begin{figure*}[t]
  \centering
  \includegraphics[width=\linewidth]{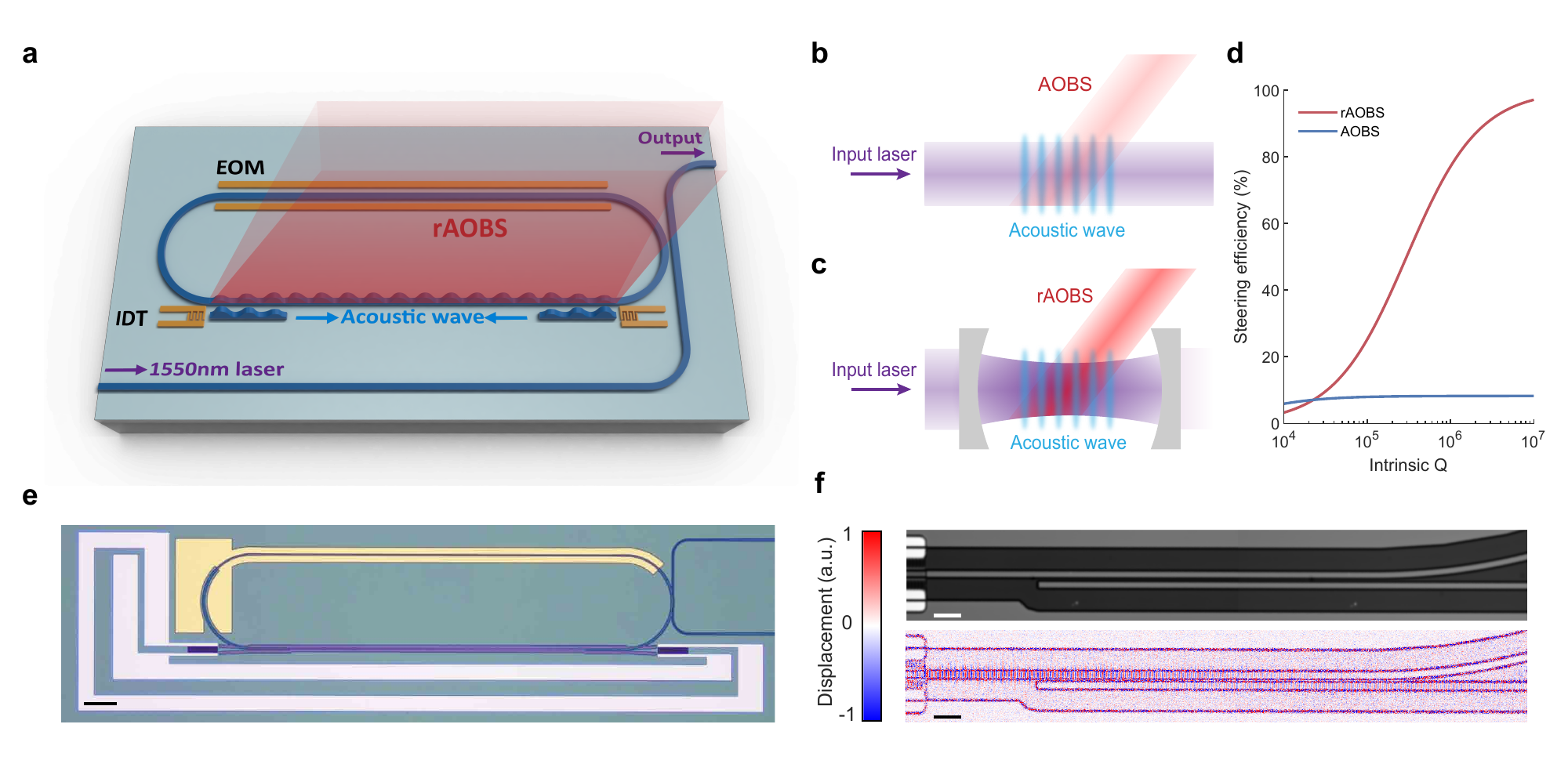}
  \caption{\textbf{Integrated resonance-enhanced acousto-optic beam steering (rAOBS).} 
  (a) Schematic diagram of the rAOBS device. A laser is coupled into and circulates within the racetrack resonator, where it undergoes resonance-enhanced acousto-optic scattering induced by acoustic wave generated by chirped IDTs.
  (b) In a single-pass AOBS, light is scattered by the acoustic wave only once, and the unscattered light is lost. 
  (c) The rAOBS recylces the light in a resonator, which is filled with acoustic waves, to resonantly enhance the acousto-optic scattering efficiency. Note that while our device utilizes a micro-ring resonator, a Fabry-Perot type of resonator is illustrated here to explain the concept. 
  (d) Simulated optimal steering efficiency of AOBS and rAOBS as functions of the intrinsic $Q$ of the resonator, assuming an acousto-optic interaction length of 1 mm and an acoustic power density of $1\ \mathrm{mW/\mu m}$.
  (e) Optical microscope image of the rAOBS device. Scale bar: $100\ \mathrm{\mu m}$.
  (f) Directional coupler of acoustic wave via evanescent coupling between two acoustic waveguides. The upper panel shows optical microscope image of coupler. The lower panel shows the surface vibrometry image of the propagating acoustic wave. Both panels are stitched from two images due to the limited FOV of the imaging system. Scale bar: $10\ \mathrm{\mu m}$.
  }\label{fig1}
\end{figure*}

Our previous demonstrations of integrated AOBS on the thin-film lithium niobate (TFLN) platform have underscored its strength for steering light from an integrated photonic chip to the free-space~\cite{li2023frequency, lin2025optical}. The principle of AOBS has been discussed in depth in~\cite{li2023frequency, lin2025optical}. The achieved optical efficiency, however, has been limited to a few percent due to the limited acousto-optic (AO) interaction length, constrained by the propagation loss of the acoustic wave and the device's single-pass construction. Further improvement in optical efficiency is imperative, especially for applications requiring high signal-to-noise ratios or operating under strict power constraints. Here, we leverage the high-quality photonic circuits realized on the TFLN platform to resonantly enhance the acousto-optical interaction by combining a high-$Q$ optical resonator with AOBS (Fig.~\ref{fig1}a). This resonance-enhanced AOBS (rAOBS) scheme significantly improves optical efficiency by extending the effective AO interaction length by more than an order of magnitude through multiple round-trip circulation of light within the resonator. In addition, the integrated LN optical resonator enables frequency chirping via electro-optic modulation, allowing the demonstration of frequency-modulated continuous-wave (FMCW) LiDAR on a compact photonic chip.

\section{Results}
\subsection*{Device design}
Our previous results~\cite{lin2025optical} have shown that, in TFLN, the propagation loss of the guided optical mode is significantly lower than that of the acoustic wave. Therefore, in AOBS, the portion of light that has not been scattered by the acoustic wave into the free space remains guided in the TFLN layer (Fig.~\ref{fig1}b). This suggests that an effective strategy for improving optical efficiency is to recycle the optical field by directing it back into the AO interaction region. This approach converts AOBS from a single-pass process into a multi-pass interaction. Optical recycling can be implemented using an optical resonator (Fig.~\ref{fig1}c), where the input laser is coupled to a resonator with the acoustic wave propagating through. When the input laser is tuned to resonance, the intra-resonator light will build up and the AO interaction will be resonantly enhanced. Fig.~\ref{fig1}d  shows the simulated steering efficiency of single-pass AOBS and resonance-enhanced AOBS for various optical Q factors, assuming an AO interaction length of 1 mm and an acoustic power density of $1\ \mathrm{mW/\mu m}$, defined as the acoustic power per unit waveguide width. For low intrinsic $Q$, the intra-resonator build-up factor is below unity, and no effective enhancement is obtained. As $Q$ increases, the optical power build-up inside the resonator becomes more pronounced. As a result, the steering efficiency of rAOBS significantly exceeds that of the single-pass configuration and eventually saturates at $100\%$ when AO scattering becomes the dominant loss channel of the resonator.

To achieve resonance-enhanced AOBS, we integrate an optical resonator with acoustic waveguides and transducers on the TFLN platform, as schematically shown in Fig.~\ref{fig1}a. The device is fabricated on a lithium niobate on insulator (LNOI) substrate with a 300-nm-thick layer of X-cut lithium niobate (LN) on a 2-$\mathrm{\mu m}$-thick layer of buried oxide. The device consists of an optical racetrack resonator coupled to a bus waveguide, two acoustic waveguides, and two broadband interdigital transducers (IDTs) that generate acoustic waves. Both the optical and acoustic waveguides are formed by partially etching the LN layer by $150\ \mathrm{nm}$ to form a rib structure. The optical and acoustic waveguides have identical width of $3\ \mathrm{\mu m}$ to guide the TE$_0$ optical mode and the Rayleigh acoustic mode, respectively. The racetrack resonator also has a pair of electro-optic (EO) electrodes, which are used to tune the resonant wavelength via LN's Pockels effect. Fig.~\ref{fig1}e shows the optical microscope image of the fabricated device. Compared to our previous devices that used slab modes of acoustic and optical waves~\cite{li2023frequency, lin2025optical}, the present design confines both acoustic and optical waves in rib waveguides, improving design flexibility and enhancing the AO interaction. 

To couple acoustic waves between two waveguides, directional couplers of acoustic wave can be designed by using evanescent coupling, in a way similar to optical directional couplers. Fig.~\ref{fig1}f shows a pair of coupled acoustic waveguides. The propagation and coupling of acoustic waves can be directly imaged using a spectral interferometry-based surface vibrometry technique~\cite{thomas2025spectral}. The vibrometry images in Fig.~\ref{fig1}f clearly shows nearly complete transfer of acoustic wave from the top waveguide to the bottom waveguide with a critical coupling length of $125\ \mathrm{\mu m}$ in the acoustic frequency range of 1.63-1.86 GHz. We use such a directional coupler to couple the acoustic waves into the racetrack resonator.The gap between the acoustic and optical waveguides is $1\ \mathrm{\mu m}$. Since the optical mode is more spatially confined than the acoustic mode in the waveguides, this separation induces negligible optical loss while preserving efficient acoustic coupling. When both IDTs are driven, the co-propagating and counter-propagating acoustic waves with respect to the optical mode generate the Stokes and anti-Stokes sidebands, respectively, through separate phase-matched traveling-wave scattering processes, thereby increasing the total steering efficiency compared with a single-IDT configuration.


\begin{figure*}[t]
  \centering
  \includegraphics[width=\linewidth]{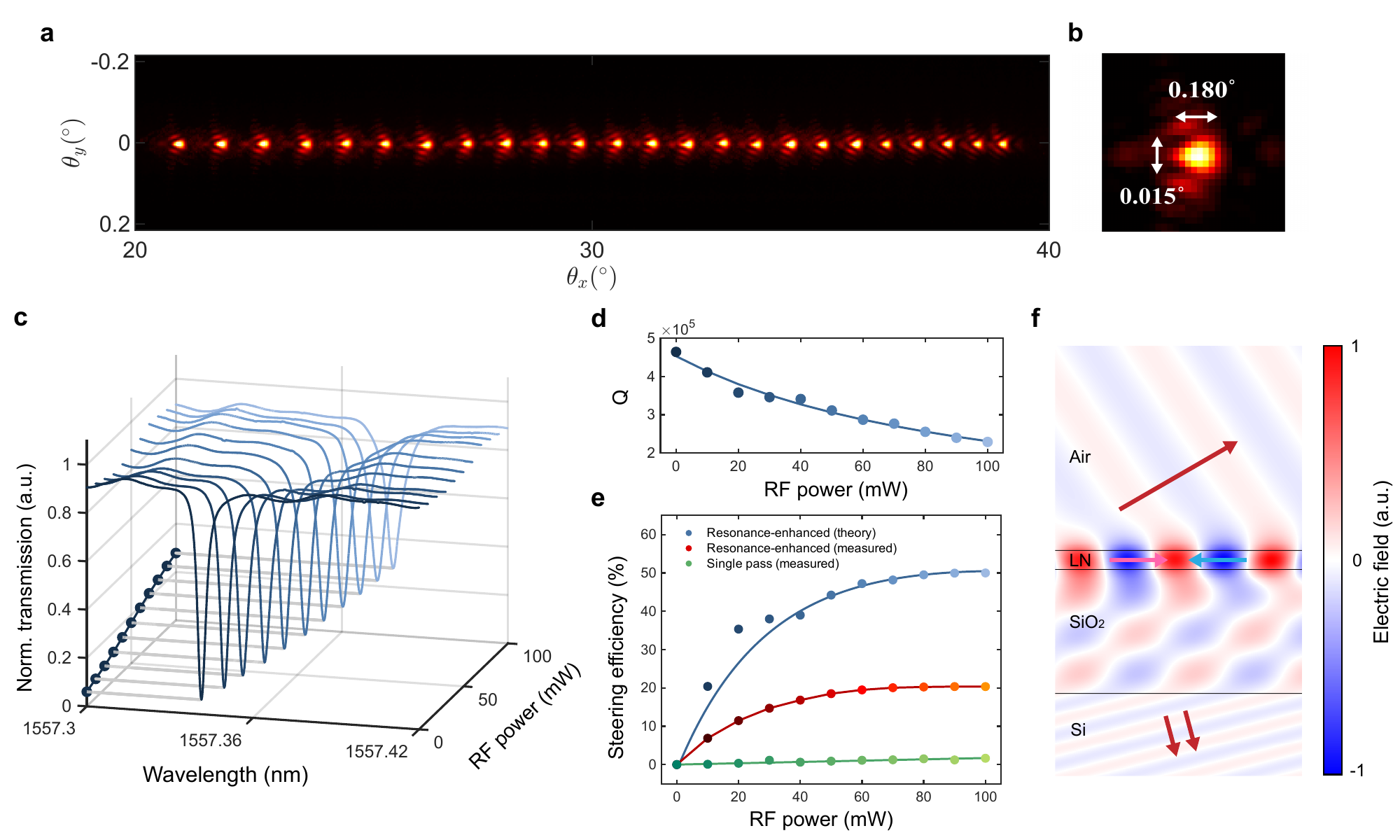}
  \caption{\textbf{Resonance-enhanced acousto-optic scattering.}
    (a) Superimposed image of the steered beams at the focal plane when the acoustic frequencies is swept from $1.63$ to $1.86\ \mathrm{GHz}$. The images were acquired at two different imaging center angles, $25^\circ$ and $35^\circ$, and stitched together to reduce the off-axis aberration caused by the finite NA of the imaging objective. Each beam profile was normalized to the same maximum pixel value before superposition.
    (b) Magnified image of the beam profile when the acoustic frequency is $1.70\ \mathrm{GHz}$. The measured angular divergence is $0.180^\circ$ along the scan direction and $0.030^\circ$ perpendicular to the scan direction. 
    (c) Evolution of the transmission spectra of the optical resonator when the RF powers used to excite the acoustic wave is increased gradually from $0$ to $100\ \mathrm{mW}$ at a fixed frequency of $1.70\ \mathrm{GHz}$. The minimum transmission points of each spectrum are projected onto the transmission-RF power plane, illustrating the progressive reduction in resonance extinction ratio with increasing RF power.
    (d) Measured loaded quality factor of the optical resonator as a function of the RF input power. The line shows a fit to the data.
    (e) Theoretical on-resonance steering efficiency calculated from panel (d), and experimentally measured on-resonance and single-pass steering efficiencies as functions of RF input power. Blue and green lines represent fits to the corresponding data, and the red line is a guide to the eye.
    (f) Simulated electric field distribution of the guided optical wave and the corresponding scattered field induced by the AOBS. Purple, blue, and red arrows denote the guided optical wave, acoustic wave, and the scattered fields, respectively. Beam steering occurs both in air and within the silicon substrate.
}
\label{fig2}
\end{figure*}

\subsection*{Resonance enhanced acousto-optic scattering}

The AO steering angle is controlled by the frequency of the acoustic wave. By tuning the acoustic frequency within the IDT bandwidth, we scanned the beam across the field of view (FOV) and measured its profile in the momentum space. Fig.~\ref{fig2}a shows the superimposed images of the steered beam captured at the focal plane of the camera system (see Supplementary Information~\ref{smsec:k_space} for optical setup). The steering angle with respect to the chip surface is centered around $30^\circ$, with a FOV of approximately $18^\circ$, limited by the RF bandwidth of the chirped IDTs. Beams from small to large steering angles correspond to acoustic frequencies ranging from 1.63 to 1.86 GHz in 10 MHz increments. As shown in Fig.~\ref{fig2}b, a Gaussian fit to the beam profile at the center of the FOV yields a full width at half maximum (FWHM) angular divergence of $0.180^\circ$ in the horizontal direction (the steering direction), which agrees well with the diffraction limit $0.886 \lambda/(L\cdot\mathrm{cos}(\varphi))=0.157^\circ$\cite{hsu2020review}, where $\lambda = 1550\ \mathrm{nm}$ is the optical wavelength, $L = 1\mathrm{mm}$ is the AO interaction length, and $\varphi=60^\circ$ is the steering angle relative to the chip-surface normal. The angular divergence perpendicular to steering is $0.015^\circ$, limited by the 5-mm-wide optical aperture of the slit filter placed after the back pupil of the objective.

The resonance-enhanced AO scattering effect can be clearly observed from the evolution of the optical resonance with increasing acoustic power. From the optical resonator's perspective, AO scattering introduces an additional loss channel to free space, suggesting that the scattering efficiency can be quantified by measuring the change in the resonator's $Q$ factor. Fig.~\ref{fig2}c shows the measured optical transmission spectra of the resonator when the RF input power used to excite the acoustic wave is progressively increased at a fixed frequency of 1.70 GHz. We observe that the loaded $Q$ gradually decreases from $4.7 \times 10^5$ to $2.1 \times 10^5$ when the RF power is increased from 0 to 100 mW  (Fig.~\ref{fig2}d, the solid line indicates a fit to the measured $Q$). In addition, the resonance red-shifts, which can be attributed to the heating effect. The decreasing $Q$ indicates an increasing strength of AO scattering. Theoretically, the resonance-enhanced AO scattering efficiency is given by
\begin{equation}
    \eta_{\mathrm{single}} =2\pi(\mathcal{F}^{-1}-\mathcal{F}_0^{-1}) \label{eq_single}
\end{equation}
\begin{equation}
    \eta_{\mathrm{rAOBS}} = B\cdot\eta_{\mathrm{single}} = \frac{\mathcal{F}}{\pi}(1-t)\eta_{\mathrm{single}}\label{eq_multi}
\end{equation}
where $\eta_{\mathrm{rAOBS}}$ and $\eta_{\mathrm{single}}$ are the on-resonance and single-pass steering efficiencies, respectively. $B$ is the build-up factor, $\mathcal{F}$ and $\mathcal{F}_0$ are the finesse with and without acoustic excitation, respectively, and $t$ is the cavity's complex transmission at resonance (see Supplementary Information~\ref{smsec:build_up_rAOBS} for derivation). At 100 mW RF input power, $\eta_{\mathrm{single}}$ and $B$  are $5.7\%$ and 8.8, respectively, therefore the theoretical on-resonance steering efficiency is $50\%$. The on-resonance steering efficiencies at various RF driving powers are calculated based on the results in Fig.~\ref{fig2}d using the same method, and plotted in Fig.~\ref{fig2}e.

Alternatively, the AO scattering efficiencies can be determined by directly measuring the power of the scattered light with an optical power meter (Thorlabs S122C) placed above the chip. To compare with rAOBS, we also fabricated and measured a single-pass device by replacing the racetrack resonator with a straight waveguide. Fig.~\ref{fig2}e shows the measured AO scattering efficiency by dividing the power of the scattered light to the optical power in the input waveguide. The efficiency of the single-pass device shows a linear dependence on the RF power, reaching $2\%$ at 100 mW according to a linear fit. In contrast, the efficiency of the rAOBS device shows a sublinear dependence on the RF power and saturates at about $20\%$ with 100 mW RF power. We note that, due to the poor impedance matching of the broadband IDTs, most of the input RF power is reflected, while less than $10\%$ is absorbed by the IDTs and converted into acoustic power. With a properly designed impedance-matching network, the input RF power required to generate the same acoustic power could be reduced to approximately 10 mW.

Comparing the two cases, the optical resonance enhances the AO scattering efficiency more than 30 times before saturation and approximately 10 times after saturation. This saturation behavior can be attributed to the change of coupling from critical-coupling regime to the under-coupling regime when the acousto-optic scattering strength increases. Using the resonator's intrinsic dissipation rate $\kappa_i$, external dissipation rate $\kappa_e$, and the AO scattering dissipation rate $\kappa_{\mathrm{AO}}$, Eq.~\ref{eq_multi} can be expressed as (see Supplementary Information~\ref{smsec:build_up_rAOBS} for derivation)
\begin{equation}
    \eta_{\mathrm{rAOBS}} = \frac{4\kappa_e\kappa_{\mathrm{AO}}}{(\kappa_i+\kappa_e+\kappa_{\mathrm{AO}})^2}
\end{equation}
Therefore, for a resonator that is critically coupled ($\kappa_i=\kappa_e$), the maximal $\eta_{\mathrm{rAOBS}}$ will be $50\%$.

The directly measured efficiency is 2.5 times lower than the values determined from the resonator $Q$ factor. This discrepancy can be explained by the portion of light that is scattered downward into the substrate, therefore, not collected by the power meter above the chip. In contrast, the decrease of the $Q$ factor is induced by the total AO scattering, regardless of the scattering direction. This is clearly revealed by the simulation result in Fig.~\ref{fig2}f, which shows that the downward scattering is more efficient than the upward one (see Supplementary Information~\ref{smsec:theory}). Although silicon is transparent at the operating telecom wavelength, direct collection of this downward-scattered light is not accessible in the current setup because the experiment does not provide backside optical access and the backside of the silicon substrate is not polished for efficient light collection. This otherwise lost power could potentially be harnessed in future designs by lowering the acoustic frequency to achieve substrate-only beam steering, together with backside optical access, a polished substrate, and dedicated out-coupling structure such as gratings or prism couplers for efficient light collection.

The optical cavity and steering efficiency characterizations described above are performed at an acoustic frequency of 1.70 GHz. The highest steering efficiency achieved within the demonstrated FOV is $26\%$, and a more comprehensive characterization across the full IDT bandwidth is provided in Supplementary Information~\ref{smsec:all_freq_measurement}.

\begin{figure*}[t]
  \centering
  \includegraphics[width=\linewidth]{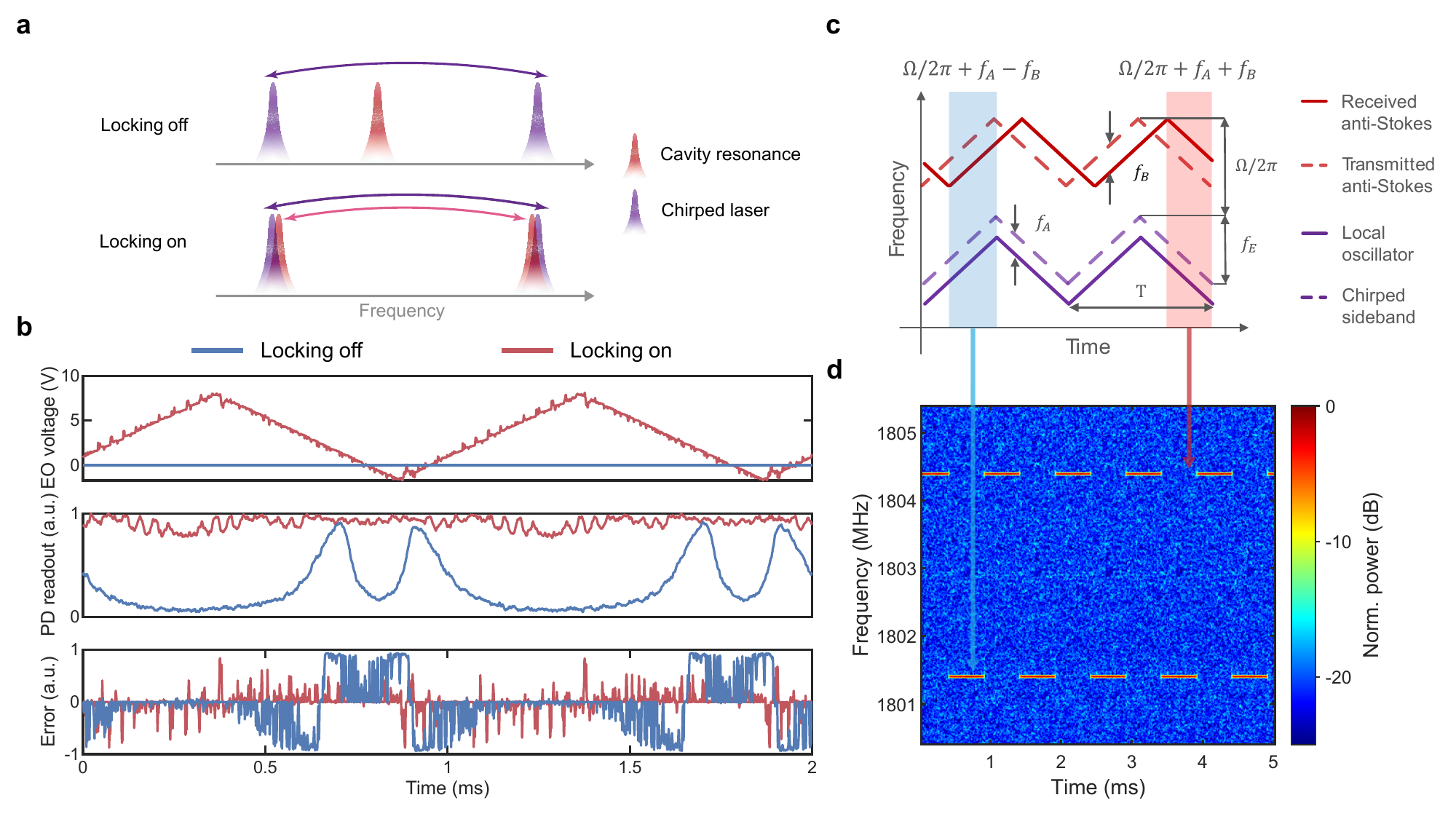}
  \caption{\textbf{Electro-optic resonance locking to a chirped laser source.} 
  (a) Illustration of electro-optic resonance locking. When the locking is off, the chirped laser sideband shifts alternately in the frequency domain, while the resonance remains fixed. When locking is on, the resonance follows the chirped sideband, resulting in enhanced beam steering.
  (b) Measured EO drive voltage from the PID controller, free-space beam power measured by a PD, and the error signal detected by the PID controller when locking is turned on and off. 
  (c) Schematic of the real-time frequencies of the lights involved in the heterodyne measurement between the received anti-Stokes steered beam and a local oscillator generated by downshifting the frequency of the chirped laser sideband using an acousto-optic frequency shifter (AOFS). The Stokes light is omitted for clarity. $\Omega$, $f_A$, $f_B$, $f_E$ and $T$ denote the acoustic angular frequency, AOFS frequency, frequency shift due to propagation delay, chirp excursion, and chirp period, respectively.
  (d) Measured real-time beat note from the signal recorded by the OSC. The diagram is generated using short-time Fourier transform.}\label{fig3}
\end{figure*}

\subsection*{Electro-optic resonance locking to a chirped laser}

A long-sought goal for the LiDAR technology is coherent FMCW ranging, which has many advantages over the incoherent time-of-flight (TOF) scheme. FMCW requires the optical system to simultaneously steer the beam and chirp the laser frequency. The ranging depth resolution $\delta d$ is set by the chirp frequency excursion $f_E$, given by  $\delta d=c/2f_E$, where $c$ is the speed of light. Therefore, to achieve a depth resolution of centimeter, $f_E$ needs to be as high as 15 GHz. The latter poses a challenge to high-$Q$ optical resonators. As depicted in Fig.~\ref{fig3}a, when the laser frequency is chirped, the detuning between the resonator resonance and the chirped laser causes substantial power fluctuations in the steered beam. To address this challenge, we utilize LN's electro-optic property by applying an external tuning voltage to the EO electrodes integrated in the resonator to dynamically lock its resonance frequency to the chirped input laser.

To lock the optical resonator to a chirped laser, we apply a weak RF signal at $20\ \mathrm{MHz}$ to the EO electrodes to dither the resonator's resonance frequency, which results in modulation of the laser transmission through the resonator. The transmitted light is detected by a photodetector, and the output signal is demodulated and used as the error signal (see Supplementary Information~\ref{smsec:slope} for details), which provides a negative feedback to a proportional-integral-derivative (PID) controller. The output of the PID controller is applied to the EO electrodes to complete the feedback loop and lock the resonator to the laser. The bandwidth of the feedback system is 10 kHz. For detailed resonance-locking setup and procedures, see Supplementary Information~\ref{smsec:locking} and~\ref{smsec:pid}. The chirped laser source is generated by modulating a CW laser with an electro-optic modulator (EOM) to produce sidebands with tunable frequencies. The EOM is driven by an amplified RF signal with a sawtooth-frequency chirped waveform, generated by an arbitrary waveform generator (AWG). We select the upper sideband as the chirped laser source, whose detuning is swept from $7$ to $12\ \mathrm{GHz}$ (chirp excursion $f_E = 5\ \mathrm{GHz}$) with a chirp period of  $T=1\ \mathrm{ms}$.

Fig.~\ref{fig3}b shows the output of the PID controller, the free-space beam power measured by a photodetector, and the error signal detected by the PID controller (see Supplementary Information~\ref{smsec:locking} for detailed experimental setup). When the locking is turned off and the laser frequency is chirped in 5 GHz range, the power of the steered optical beam (blue, middle panel) varies over ten times with chirping. When the locking is turned on, the EO drive voltage (red, top panel) supplied by the PID controller faithfully tracks the sawtooth chirping of the laser. The locking results in a substantially more stable optical power in the steered beam (red, middle panel). The periodic feature observed in the error signal is likely associated with chirp-synchronous residual tracking error in the resonance-locking loop. During laser frequency tracking, the cavity resonance is tuned to follow the periodically chirped optical sideband, while the lock-in detector, PID controller, EOM driver, and feedback electronics all have finite bandwidth and can introduce phase delay. This residual error is expected to be more pronounced near the turning points of the sawtooth chirp, where the tuning trajectory reverses rapidly and the feedback loop may exhibit small overshoot or undershoot. Since the error remains bounded when the lock is on, this feature is attributed to residual tracking error rather than loss of locking.

To confirm the coherence of the chirped laser steered by the device, we perform a heterodyne measurement between the steered light (collected by a collimator and delayed through a fiber delay line) and the local oscillator (LO), which is the essential process of FMCW ranging. In FMCW LiDAR, a frequency-chirped optical signal is transmitted toward the target and reflected back with a propagation delay $\Delta t=2d/c$ proportional to the target distance $d$. The reflected signal is then mixed with a replica of the chirped laser serving as a local oscillator, producing a beat frequency $f_B=2f_E\Delta t/T$ through heterodyne detection, where $f_E$ and $T$ are the frequency chirp excursion and period. Since the frequency difference between the two signals is proportional to the round-trip delay, the target distance can be directly extracted from $f_B$. When implementing FMCW LiDAR with rAOBS, the additional frequency shift introduced by AO scattering must also be taken into account. Specifically, co-propagating and counter-propagating acoustic waves with respect to the optical propagation direction generate Stokes and anti-Stokes sidebands, respectively, resulting in frequency-shifted steered beams. We focus on the anti-Stokes sideband in the following discussion. As shown Fig.~\ref{fig3}c, the heterodyne measurement produces two beating frequencies of $\Omega/2\pi \pm f_B$, where $\Omega$ is the angular frequency of the acoustic wave. To differentiate Stokes and anti-Stokes, the LO frequency is down-shifted by $f_A = 102.9\ \mathrm{MHz}$ using an acousto-optic frequency shifter (AOFS)~\cite{liu2019electromechanical} (see Supplementary Information~\ref{smsec:locking} for detailed experimental setup). This scheme results in two beat notes in the received signal:  $\Omega/2\pi + f_A \pm f_B$ for the anti-Stokes sideband ($\Omega/2\pi - f_A \pm f_B$ for the Stokes sideband). Fig.~\ref{fig3}d  illustrates the real-time frequencies of the anti-Stokes sideband and the shifted LO, where $\Omega/2\pi = 1.70\ \mathrm{GHz}$, and $f_B\approx1.5\ \mathrm{MHz}$ (corresponding to $\sim45\ \mathrm{m}$ delay length, including both free-space and fiber propagation paths.). The heterodyne beat note is recorded with an oscilloscope and analyzed with short-time Fourier transform (STFT) to capture its real-time spectrum, as shown in Fig.~\ref{fig3}d. The spectrograph clearly shows two alternating frequency components at $\Omega/2\pi + f_A \pm f_B$ in agreement with the analysis in Fig.~\ref{fig3}c.

\begin{figure*}[t]
  \centering
  \includegraphics[width=\linewidth]{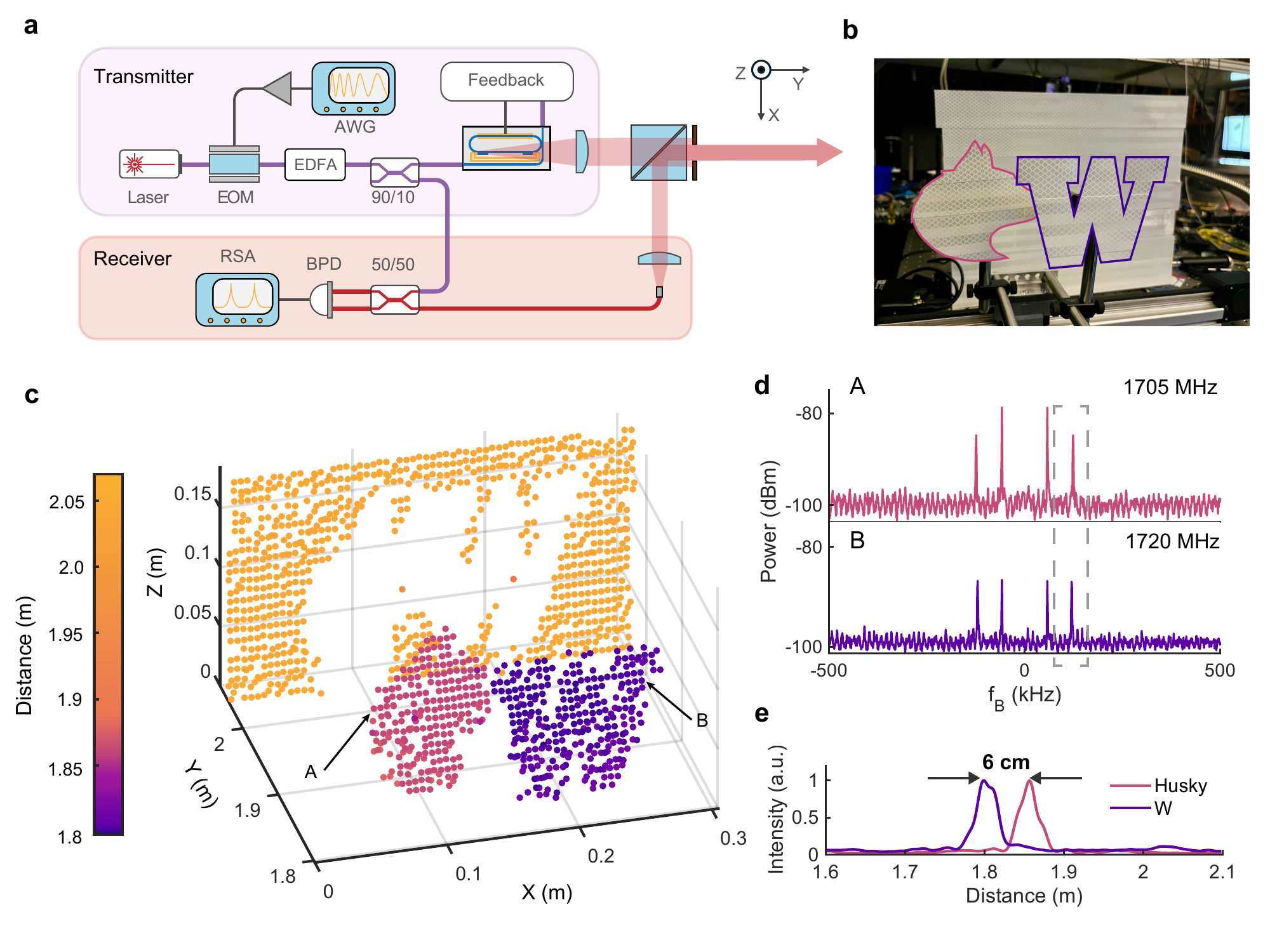}
  \caption{\textbf{FMCW LiDAR demonstration.}
    (a) Schematic diagram of the measurement system for resonance locking and FMCW LiDAR demonstration. The detailed diagram of the feedback block can be found in the Supplementary Information~\ref{smsec:locking}. AWG: arbitrary waveform generator; EOM: electro-optic modulator; EDFA: erbium-doped fiber amplifier; RSA, real-time signal analyzer; BPD: balanced photodetector; 
    (b) Photograph of the LiDAR target. The Husky and W icons are outlined with colored lines.
    (c) Point cloud of the target captured by the LiDAR.
    (d) Representative heterodyne beat-note spectra of signals reflected from the Husky and W (points A and B in panel (c)). Each spectrum is centered around its respective acoustic frequency, 1705 MHz for point A and 1720 MHz for point B. The inner and outer pairs of peaks correspond to reflections from the distance reference reflector and the LiDAR target, respectively.
    (e) Zoomed-in view of two FMCW signals calculated from the data in the dash box in (d).}\label{fig4}
\end{figure*}

\subsection*{FMCW LiDAR demonstration}
By leveraging the high steering efficiency and electro-optic resonance locking of the rAOBS chip, we demonstrate a proof-of-concept FMCW LiDAR. The experimental setup is shown in Fig.~\ref{fig4}a and Fig.~\ref{fig4}b, and comprises three parts: transmitter, receiver, and LiDAR target. At the transmitter, a chirped laser sideband with a 5 GHz excursion and 1 ms chirp period is generated, amplified by an erbium-doped fiber amplifier (EDFA), and sent to the rAOBS chip through a 90/10 fiber coupler. The rAOBS chip is locked on-resonance to the chirped laser sideband. The laser beam steered by the rAOBS in to the free-space is directed to a beam splitter and illuminates a LiDAR target located approximately 2 m away. At the receiver, the reflected light is collected through the beam splitter and a collimator, and fed into a fiber for heterodyne detection. The resulting beat-note spectrum is recorded using a real-time signal analyzer (RSA) to extract the target distance.

The targets are cutouts made of reflective film, as shown in Fig.~\ref{fig4}b. In this demonstration, we use the rAOBS device to scan the beam along the vertical direction, while mechanically translating the target to achieve scanning along the horizontal direction. In future LiDAR systems, multiple rAOBS devices in an array along the horizontal direction can be used to enable simultaneous multi-line scanning, thereby eliminating the need for any mechanical actuation.

The measured point cloud is shown in Fig.~\ref{fig4}c, which shows clearly-resolved shapes of the targets. Distinct separation of different components along the depth (Y axis) and high angular resolution along the Z axis are observed. With $f_E=5$ GHz, the ranging depth resolution $\delta d$ is 3 cm. Two representative heterodyne beat-note spectra recorded by the RSA for signals reflected from the Husky and the W (point A and B in Fig.~\ref{fig4}c) are shown in Fig.~\ref{fig4}d. The heterodyne beat notes of A  and B are centered around 1705 MHz and 1720 MHz, respectively, as the steered light is shifted by the acoustic frequency compared to the laser source due to AO scattering. The extra frequency shifts ($f_\mathrm{B}$) are due to the propagation delay of the steered beams in free space and fibers. Each spectrum exhibits two pairs of peaks: the inner pair of peaks originates from a distance reference target (not shown in the schematic), while the outer pair corresponds to reflections from the targets. A zoomed-in view of the FMCW ranging results calculated from the data in the dash box of Fig.~\ref{fig4}d is shown in Fig.~\ref{fig4}e, revealing a depth difference of 6 cm between the W and Husky features. The FWHM of each beat-note peak corresponds to a ranging resolution of 3 cm, consistent with theoretical expectation.

\section{Discussion}

In this work, we demonstrate enhancement of acousto-optic beam steering by using an integrated optical resonator on the TFLN paltform, achieving up to $26\%$ optical efficiency and a field of view of $18^\circ$. Using an electro-optic modulator integrated on the resonator, we implement a resonance-locking scheme that synchronizes the optical resonator to an externally frequency-chirped laser with GHz excursion, and demonstrate FMCW LiDAR with centimeter resolution. Further improvements are expected through engineering of the material platform to reduce acoustic wave propagation loss~\cite{lin2026experimental}. In addition, we find that acousto-optic scattering into the substrate is significantly more efficient than scattering into air, suggesting that the use of a transparent substrate could further enhance performance. This device platform combines LN’s piezoelectric and electro-optic properties to enable multifunctional operation. The combination of high optical efficiency, agile angular steering, and electro-optic modulation capabilities makes this platform well-suited for applications including FMCW LiDAR and free-space optical communication, with further device- and system-level engineering effort to improve scalability, compactness, and power consumption.

\section*{Methods} 
\subsection*{Device fabrication}
The rAOBS is fabricated on X-cut lithium niobate on insulator (LNOI) wafers with $300\ \mathrm{nm}$ thick LN layer, $2\ \mathrm{\mu m}$ thick thermal oxide layer and $525\ \mathrm{\mu m}$ thick silicon substrate (from NanoLN Inc.). The optical waveguides are patterned with electron-beam lithography (EBL) and then etched for $150\ \mathrm{nm}$ using ion beam etcher (IBE). After stripping the resist, piranha and standard cleaning solution are used to remove resist residue and redeposition. The IDT and EO electrode were patterned with electron-beam lithography and followed by a lift-off process of $1\ \mathrm{nm}$ Ti - $180\ \mathrm{nm}$ Al, and $2\ \mathrm{nm}$ Ti - $100\ \mathrm{nm}$ Au, respectively. 

\section*{Acknowledgements}
This work is supported by the National Science Foundation (Award No. ITE-2134345 and OSI-2326746) and the DARPA MTO SOAR program (Award No. HR0011363032). Part of this work was conducted at the Washington Nanofabrication Facility and Molecular Analysis Facility, a National Nanotechnology Coordinated Infrastructure (NNCI) site at the University of Washington with partial support from the National Science Foundation via award nos. NNCI-2025489. Development of the optical vibrometer was partially supported by the DARPA DSO OPTIM program (Award No. HR00112320031).

\section*{Author Contributions}
Y.Y. and Q.L. conceived the work, with valuable discussion from S.F., B.L. and M.L.. Y.Y. and Q.L. fabricated the device with assistance from Z.X., J.J. and L.S.. Y.Y. and Q.L. performed the experiment with assistance from S.F. and Y.Z.. J.G.T. and Y.Z. performed the acoustic wave vibrometry imaging. Y.Y. and M.L. prepared the manuscript with discussion and input from all authors. B.L. and M.L. supervised the work.
\section*{Competing interests}

\bibliography{sample_rev_1}{}
\bibliographystyle{ieeetr}








\clearpage

\appendix
\onecolumngrid









\newpage

\renewcommand{\thefigure}{S\arabic{figure}}
\setcounter{figure}{0}
\renewcommand{\thetable}{S\arabic{table}}
\setcounter{table}{0}
\renewcommand{\thesection}{S\arabic{section}}
\setcounter{section}{0}
\renewcommand{\theequation}{\thesection.\arabic{equation}}
\setcounter{equation}{0}                

\section{Perturbation theory for acousto-optic scattering}\label{smsec:theory}
The wave equation for AO scattering process is:

\begin{equation}
    [\nabla^2 + \omega^2\mu_0\varepsilon_0(\varepsilon_r(y,z) + \Delta\varepsilon_r(x,y,z))]\mathbf{E}(x,y,z) = 0\label{smfunc:eq_perturb}
\end{equation}
where the propagation direction is along the x-axis, $\varepsilon_r(y,z)$ denotes the relative permittivity of the unperturbed photonic waveguide, and $\Delta \varepsilon_r(x,y,z)$ represents the perturbation induced by the acoustic wave.

The zeroth order solution of Eq.~\ref{smfunc:eq_perturb} corresponds to the waveguide mode, and its equation and solution are given by

\begin{equation}
    [\nabla^2 + \omega^2\mu_0\varepsilon_0\varepsilon_r(y,z) ]\mathbf{E}_0(x,y,z) = 0\label{smfunc:eq_0}
\end{equation}
\begin{equation}
    \mathbf{E}_0(x,y,z) = A_0\mathbf{E}_{\mathrm{wg}}(y,z)\exp(ik_{\mathrm{wg}}x - i\omega t) + c.c.\label{smfunc:sol_0}
\end{equation}
where $A_0$ is a dimensionless complex amplitude, $\mathbf{E}_{\mathrm{wg}}$ denotes the mode of interest (for example, TE$_0$ in our case), $\omega$ is the angular frequency, $k_\mathrm{wg} = n_\mathrm{eff}k_0$ is the wavenumber of the guided mode, and $c.c.$ stands for complex conjugate. $\mathbf{E}_0(x,y,z)$ should include $t$ as an argument; however, we omit it here and in all subsequent expressions for the electric field, as the equations used in the following analysis are time-independent.

For the first order perturbation, the waveguide mode is scattered into free space, where the field is unguided and continuously distributed, making it difficult to describe using traditional coupled-mode theory. An alternative approach is to treat the zeroth-order solution as a non-decaying excitation and directly solve for the radiation field it generates. In reality, the waveguide field decays along its propagation direction, and consequently the radiation field it excites will differ from the ideal case. To justify this approximation, in our system the decay of the waveguide mode occurs over a spatial scale much longer than the distance required for the radiation field in free space to establish. Therefore, it is a reasonable approximation to treat the waveguide mode as a non-decaying field. More importantly, only the near field of the radiation field contributes significantly to the decay of the guided mode due to AO scattering, which is the quantity of interest in our analysis.

Extending Eq.~\ref{smfunc:eq_perturb} to first order yields the following equation:
\begin{equation}
    [\nabla^2 + \omega^2\mu_0\varepsilon_0(\varepsilon_r(y,z) + \Delta\varepsilon_r(x,y,z))](\mathbf{E}_0(x,y,z)+\mathbf{E}_1(x,y,z)) = 0\label{smfunc:eq_perturb_1}
\end{equation}

By neglecting the second-order terms and subtracting Eq.~\ref{smfunc:eq_0}, we obtain
\begin{equation}
    [\nabla^2 + \omega^2\mu_0\varepsilon_0\varepsilon_r(y,z) ]\mathbf{E}_1(x,y,z) = -\omega^2\mu_0\varepsilon_0\Delta\varepsilon_r(x,y,z)\mathbf{E}_0(x,y,z)\label{smfunc:eq_1}
\end{equation}

The acoustic-induced relative permittivity change, $\Delta\varepsilon_r(x,y,z)$, consists of two contributions: the photo-elastic term $\Delta\varepsilon_\mathrm{r,pe}$ and the moving-boundary term $\Delta\varepsilon_\mathrm{r,mb}$. The photo-elastic term describes the permittivity change due to the photo-elastic effect, and is given by:
\begin{equation}
    \Delta\varepsilon_{\mathrm{r, pe},ij} = -\frac{\varepsilon_{ik}\varepsilon_{jl}p_{klmn}S_{mn}}{\varepsilon_0}
\end{equation}
where $p_{klmn}$ and $S_{mn}$ are the photo-elastic tensor and strain tensor, respectively. The moving boundary term is simply given by the changes in $\varepsilon_r$ distribution due to geometry deformation: 
\begin{equation}
    \Delta\varepsilon_\mathrm{r, mb}(x,y,z) = \mathbf{u}\cdot\mathbf{\hat{n}}[\varepsilon_\mathrm{r1}(x,y,z)-\varepsilon_\mathrm{r2}(x,y,z)]\delta_{\partial R}
\end{equation}
where $\mathbf{u}$ is the surface displacement vector, $\mathbf{\hat{n}}$ is the surface normal vector pointing from region 1 to region 2, and $\delta_{\partial R}$ is the Dirac delta function defined on the interface $\partial R$ between regions 1 and 2.

We denote the change in relative permittivity due to Rayleigh mode with the following term:
\begin{equation}
    \Delta\varepsilon_r(x,y,z) = \Delta\varepsilon_{\mathrm{r, pe}} + \Delta\varepsilon_{\mathrm{r, mb}} = [A_\mathrm{ac}\Delta\varepsilon_r(y,z) \exp(iK x) + c.c.]/2\label{smfunc:delta_eps}
\end{equation}
Here, $K$ is the wavenumber of the acoustic wave, and $A_\mathrm{ac}$ is the dimensionless complex amplitude of acoustic wave. A term $\Omega$ representing the angular frequency of the acoustic wave should also be included; however, to simplify the analysis, we omit it and treat the acoustic perturbation as a static perturbation. The only consequence of adding $\Omega$ back into equations is that the angular frequency of the radiation field must be shifted by $\pm \Omega$, depending on whether the scattering process absorbs or emits a phonon.

By substituting Eq.~\ref{smfunc:delta_eps} into Eq.~\ref{smfunc:eq_1}, the equation can be solved to yield the following first-order radiation field:
\begin{gather}
    \mathbf{E}_1(x,y,z) = A_0A_\mathrm{ac}^\ast\mathbf{E}_\mathrm{rad}(y,z)\exp[i(k_\mathrm{wg} - K)x  - \omega t] +c.c.
\end{gather}

So far, no term has been found that accounts for the back-action of AO scattering on the guided optical field. Therefore, we extend the perturbation analysis to second order:

\begin{equation}
    [\nabla^2 + \omega^2\mu_0\varepsilon_0(\varepsilon_r(y,z) + \Delta\varepsilon_r(x,y,z))](\mathbf{E}_0(x,y,z)+\mathbf{E}_1(x,y,z) + \mathbf{E}_2(x,y,z)) = 0\label{smfunc:eq_perturb_2}
\end{equation}

By neglecting the third-order terms and subtracting Eq.~\ref{smfunc:eq_0} and Eq.~\ref{smfunc:eq_1}, we obtain

\begin{equation}
    [\nabla^2 + \omega^2\mu_0\varepsilon_0\varepsilon_r(y,z) ]\mathbf{E}_2(x,y,z) = -\omega^2\mu_0\varepsilon_0\Delta\varepsilon_r(x,y,z)\mathbf{E}_1(x,y,z)\label{smfunc:eq_2}
\end{equation}

In Eq.~\ref{smfunc:eq_2}, the right-hand side produces two terms with wavenumbers $k_{wg} - 2K$ and $k_{wg}$. Since the $k_{wg}$ term corresponds to the waveguide mode, it builds up much faster than the $k_{wg} - 2K$ term; therefore, we focus only on the $k_{wg}$ term in the analysis. Based on couple-mode theory, under slow-varying approximation, we have

\begin{equation}
    \mathbf{E}_2(x,y,z) = A_2(x)[\mathbf{E}_{\mathrm{wg}}(y,z)\exp(ik_{\mathrm{wg}}x - i\omega t) + c.c.]
\end{equation}
\begin{equation}
    \frac{\mathrm{d}A_2(x)}{\mathrm{d}x} = iA_0A_\mathrm{ac}^\ast A_\mathrm{ac}\frac{\omega^2\mu_0\varepsilon_0}{2k_\mathrm{eff}}\frac{\iint \mathrm{d}y\mathrm{d}z\mathbf{E}_\mathrm{wg}^\ast(y,z)(\Delta\varepsilon_r(y,z)/2) \mathbf{E}_\mathrm{rad}(y,z)}{\iint \mathrm{d}y\mathrm{d}z\mathbf{E}_\mathrm{wg}^\ast(y,z)\mathbf{E}_\mathrm{wg}(y,z)}
\end{equation}

We note that only the radiation field overlapping with $\mathbf{E}_\mathrm{wg}$ via $\Delta\varepsilon_r(y,z)$ affects the guided wave. Therefore, only the near field of the radiation field contributes significantly, which justifies our approximation in the first-order perturbation.

Since $\mathbf{E}_0$ and $\mathbf{E}_2$ belong to the same guided mode, their sum $A = A_0 + A_2$ corresponds to the overall field amplitude in this mode. Consequently, the total amplitude $A(x)$ obeys the following equation of motion:

\begin{equation}
    \frac{\mathrm{d}A(x)}{\mathrm{d}x} = iA(x)|A_\mathrm{ac}|^2\frac{\omega^2\mu_0\varepsilon_0}{2k_\mathrm{eff}}\frac{\iint \mathrm{d}y\mathrm{d}z\mathbf{E}_\mathrm{wg}^\ast(y,z)(\Delta\varepsilon_r(y,z)/2) \mathbf{E}_\mathrm{rad}(y,z)}{\iint \mathrm{d}y\mathrm{d}z\mathbf{E}_\mathrm{wg}^\ast(y,z)\mathbf{E}_\mathrm{wg}(y,z)}\label{smfunc:eom1}
\end{equation}

Neglecting the self-coupling terms (the imaginary part of the right-hand side of Eq.~\ref{smfunc:eom1}, which only modifies the complex angle of $A(x)$ and does not change its amplitude), we obtain the following $A(x)$:

\begin{equation}
    A(x) = A(0)\exp(-\alpha x/2)
\end{equation}
\begin{equation}
    \alpha =  |A_\mathrm{ac}|^2\mathrm{Im}\left[\frac{\omega^2\mu_0\varepsilon_0}{k_\mathrm{eff}}\frac{\iint \mathrm{d}y\mathrm{d}z\mathbf{E}_\mathrm{wg}^\ast(y,z)(\Delta\varepsilon_r(y,z)/2) \mathbf{E}_\mathrm{rad}(y,z)}{\iint \mathrm{d}y\mathrm{d}z\mathbf{E}_\mathrm{wg}^\ast(y,z)\mathbf{E}_\mathrm{wg}(y,z)}\right]
\end{equation}
and $\alpha$ is the energy decay rate of the guided optical wave in the AO scattering process.

\begin{figure}[htbp]
  \centering
  \includegraphics[width=\linewidth]{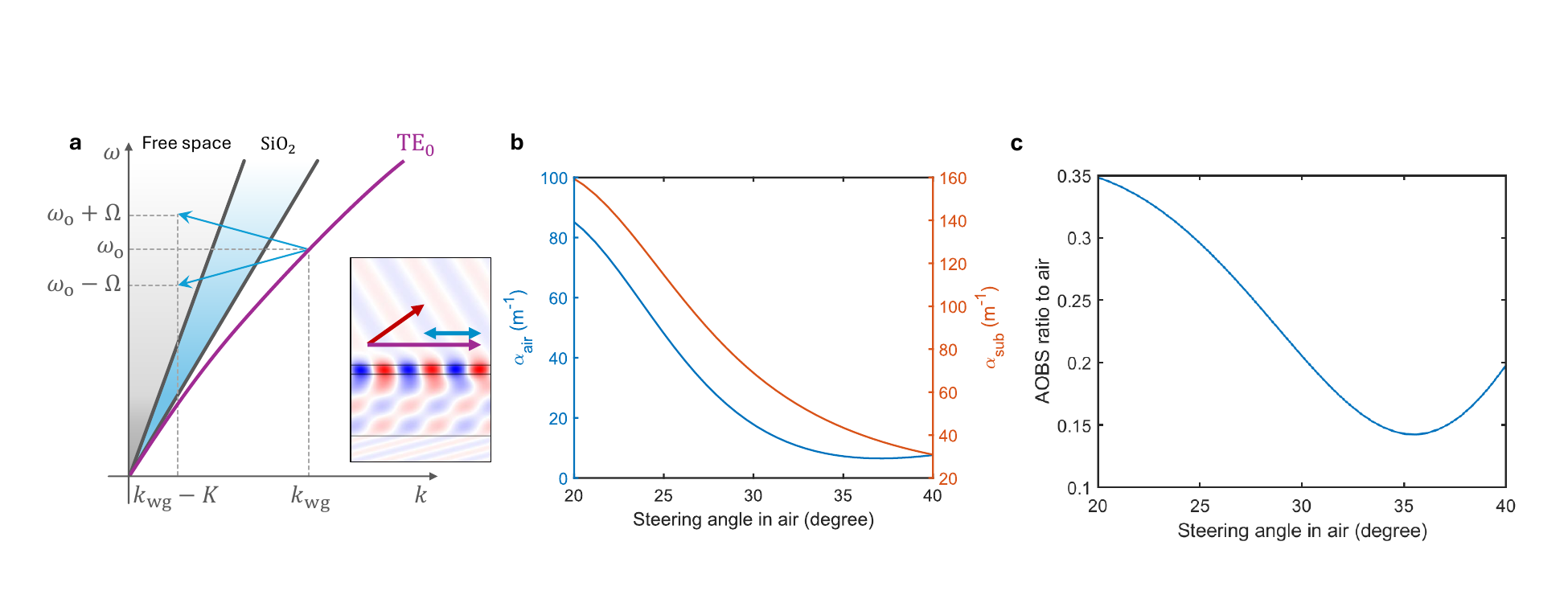}
  \caption{\textbf{Simulated acousto-optic scattering efficiency.} 
  (a) Dispersion diagram of rAOBS. acoustic waves (blue arrows) scatter the TE$_0$ light into the free-space light cone via Stokes and anti-Stokes processes. Inset shows the simulated TE$_0$ field and its scattered fields in free space and substrate. Material of layers from bottom to top are silicon, thermal oxide, x-cut lithium niobate, and air. 
  (b) The blue and orange curves correspond to $\alpha_{\mathrm{air}}$ and $\alpha_{\mathrm{sub}}$, respectively. These results are simulated under an acoustic power density of $1\ \mathrm{mW}/\mathrm{\mu m}$, meaning that a $1\ \mathrm{\mu m}$-wide waveguide carries $1\ \mathrm{mW}$ of acoustic power.
  (c) The ratio of the optical power steered into air to the total power steered into both air and the substrate, as a function of steering angle in air. This results are calculated based on (b).}\label{smfig:air_sub_eff}
\end{figure}

The AO scattering efficiency from the guided mode into free space and the substrate can also be calculated using first-order perturbation theory. Since the radiation field excited by the guided mode is obtained from the first order perturbation, the scattering efficiency is defined as the ratio between the integral of out-of-chip projection of the averaged Poynting vector generated by $E_1$ and the integral of the averaged Poynting vector of $E_0$ in the propagation direction:

\begin{equation}
    \alpha_\mathrm{air} = \frac{\int \mathrm{d}z\mathbf{\hat{y}}\cdot \bar{\mathbf{P}}_\mathrm{1,air}}{\iint \mathrm{d}y\mathrm{d}z\mathbf{\hat{x}}\cdot \bar{\mathbf{P}}_\mathrm{0}}
\end{equation}
\begin{equation}
    \alpha_\mathrm{sub} = \frac{\int \mathrm{d}z(-\mathbf{\hat{y}})\cdot \bar{\mathbf{P}}_\mathrm{1,sub}}{\iint \mathrm{d}y\mathrm{d}z\mathbf{\hat{x}}\cdot \bar{\mathbf{P}}_\mathrm{0}}
\end{equation}

Here, we assume that the surface normal vector of the chip is $\mathbf{\hat{y}}$, pointing toward the air. Based on these equations, we simulate and plot $\alpha_{\mathrm{air}}$ and $\alpha_{\mathrm{sub}}$ at an acoustic power density of $1\ \mathrm{mW}/\mathrm{\mu m}$. The dispersion diagram is shown in Fig.~\ref{smfig:air_sub_eff}a, and the inset shows the field distribution simulated based on the methods above. In Fig.~\ref{smfig:air_sub_eff}b, we show the simulated $\alpha_\mathrm{air}$ and $\alpha_\mathrm{sub}$ within the steering angle of our device. The simulation indicates that the AO scattering efficiency into the substrate is generally higher than that into the air. Fig.~\ref{smfig:air_sub_eff}c shows the ratio of the optical power steered into air to the total steered power, including contributions to both air and the substrate.The calculated ratio is lower than the experimentally measured value (approximately $40\%$ at about $26^\circ$). This discrepancy may arise from measurement uncertainties or from differences between the simulation model and the experiment, as the simulation assumes slab optical and acoustic modes, whereas the experiment uses a $3\ \mathrm{\mu m}$-wide waveguide.

\section{Derivation of the resonance build-up factor in rAOBS}\label{smsec:build_up_rAOBS}

For a cavity with internal dissipation rate $\kappa_i$, external dissipation rate $\kappa_e$, and free spectral range (FSR), its build-up factor, which represents the intracavity energy enhancement relative to the input waveguide when on resonance, is defined as
\begin{equation}
B = \frac{|E_\mathrm{cavity}|^2}{|E_\mathrm{input}|^2}
\end{equation}
where $E_\mathrm{cavity}$ and $E_\mathrm{input}$ denote the intracavity and input waveguide electric fields, respectively. 

When on resonance, the input field couples into the cavity and circulates within it (for a ring cavity). The portion of energy that couples from the input waveguide into the cavity is 
\begin{equation}
    1-\exp(-\kappa_e T_\mathrm{rt}) \approx \kappa_e T_\mathrm{rt}=\frac{\kappa_e}{\mathrm{FSR}}
\end{equation}
Here, $T_\mathrm{rt}$ is the round-trip time of the cavity, defined as the time required for light to complete one round trip, and is related to the free spectral range by $T_\mathrm{rt} = 1/\mathrm{FSR}$.

The circulating field amplitude decays exponentially as $\exp(-i/2N_\mathrm{rt})$, where $i$ denotes the number of round trips, and $N_\mathrm{rt}$ is the characteristic number of round trips over which the field energy decays to $1/e$ of its initial value. The parameter $N_\mathrm{rt}$ is given by
\begin{equation}
N_\mathrm{rt} = \frac{1}{(\kappa_i+\kappa_e)T_\mathrm{rt}} = \frac{\mathrm{FSR}}{\kappa_i + \kappa_e}
\end{equation}

Accordingly, the intracavity field builds up as a coherent sum over successive round trips, yielding
\begin{equation}
B = \frac{\kappa_e}{\mathrm{FSR}} \left| \sum_{i=0}^\infty \exp\left(-\frac{i}{2N_\mathrm{rt}}\right) \right|^2
= \frac{4\kappa_e N_\mathrm{rt}^2}{\mathrm{FSR}}
= \frac{4\kappa_e\mathrm{FSR}}{(\kappa_i+\kappa_e)^2}
\end{equation}\label{smfunc:build_up_1}

Eq.~\ref{smfunc:build_up_1} can be further simplified by introducing the cavity finesse and complex transmission. The finesse is defined as the ratio of the free spectral range to the resonance linewidth,
\begin{equation}
F = \frac{\mathrm{FSR}}{(\kappa_i + \kappa_e)/2\pi}
= \frac{2\pi\mathrm{FSR}}{\kappa_i+\kappa_e}
\end{equation}
The on-resonance complex transmission of the cavity is given by
\begin{equation}
t = \frac{\kappa_i-\kappa_e}{\kappa_i+\kappa_e}
\end{equation}
Accordingly, the build-up factor can be expressed as
\begin{equation}
B = \frac{F}{\pi}\left(1-t\right)
\end{equation}\label{smfunc:build_up_2}

For rAOBS, AO scattering can be regarded as an intrinsic dissipation channel, therefore we can replace $\kappa_i$ in Eq.~\ref{smfunc:build_up_1} with $\kappa_i + \kappa_\mathrm{AO}$:
\begin{equation}
    B = \frac{4\kappa_e \mathrm{FSR}}{(\kappa_i + \kappa_\mathrm{AO}+\kappa_e)^2}
\end{equation}
The AO scattering loss per round trip is
\begin{equation}
    \eta_\mathrm{single} = 1-\exp(-\kappa_\mathrm{AO}T_\mathrm{rt})\approx\kappa_\mathrm{AO}T_\mathrm{rt}=\frac{\kappa_\mathrm{AO}}{\mathrm{FSR}}
\end{equation}
therefore the steering efficiency of rAOBS is 
\begin{equation}
    \eta_\mathrm{rAOBS} = B\cdot\eta_\mathrm{single} = \frac{4\kappa_e\kappa_\mathrm{AO}}{(\kappa_i + \kappa_\mathrm{AO}+\kappa_e)^2}
\end{equation}

\section{Device design and simulation}\label{smsec:design}
In our design, the lithium niobate film has a thickness of $300\ \mathrm{nm}$ and is etched to a depth of $150\ \mathrm{nm}$ to define both the photonic and acoustic waveguides. The widths of the photonic and acoustic waveguides are both $3\ \mathrm{um}$, with a coupling gap of $1\ \mathrm{um}$ between them. Finite-element-method (FEM) simulations show that the coupling length required for $100\%$ power transfer of the TE$_0$ optical mode at a wavelength of $1550\ \mathrm{nm}$ is $4.23\ \mathrm{mm}$. In contrast, the coupling length required for $100\%$ energy transfer of the Rayleigh mode at $1.70\ \mathrm{GHz}$ is $125\ \mathrm{um}$.

Based on these results, we set the acoustic wave coupling length to $125\ \mathrm{um}$. Over this interaction length, the simulated power transfer of the $1550\ \mathrm{nm}$ TE$_0$ optical mode is only $0.2\%$, indicating a negligible perturbation to the optical field propagation within the resonator.

\begin{figure}[htbp]
  \centering
  \includegraphics[width=\linewidth]{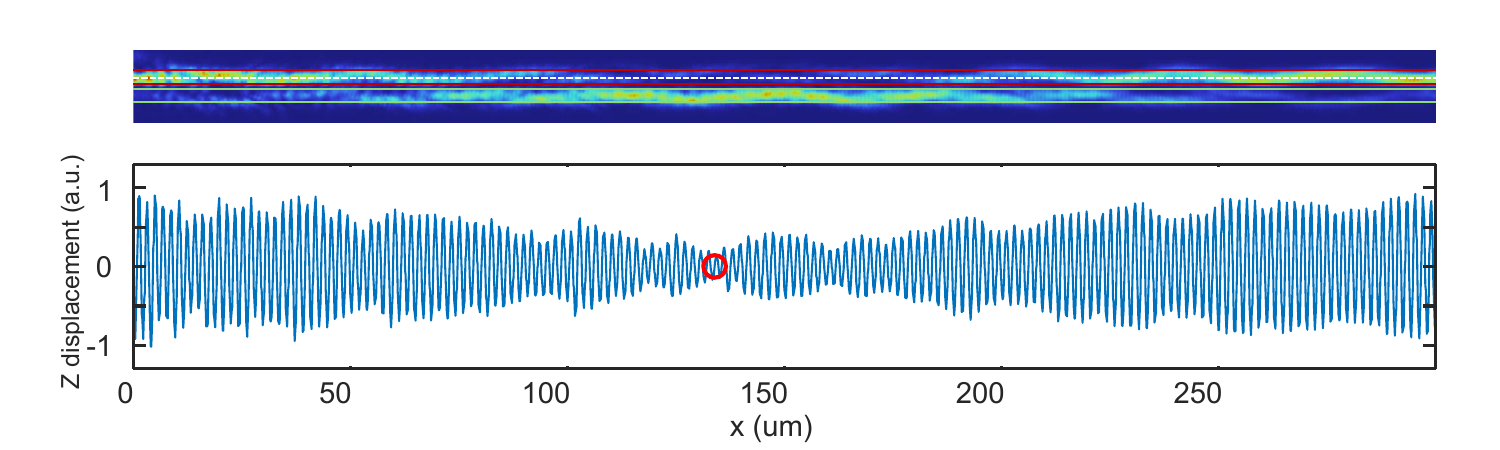}
  \caption{\textbf{FDTD simulation of evanescent coupling of acoustic wave.} The top panel shows a top-down view of two acoustic waveguides, with the waveguide edges indicated by red and green lines. An incident acoustic wave is injected from the left end of the top waveguide. The bottom panel shows the out-of-plane displacement along the center of the top waveguide (marked by the white dashed line). The position corresponding to the minimum displacement is highlighted with a red circle, located at $135\ \mathrm{um}$.}\label{smfig:SAW_fdtd}
\end{figure}

We further verified the acoustic wave coupling length using frequency-domain simulations in COMSOL Multiphysics, and the results are shown in Fig.~\ref{smfig:SAW_fdtd}. The simulated spatial evolution of the acoustic energy along the waveguide direction agrees with the FEM results. The coupling length obtained from the simulation is $135\ \mathrm{um}$, in close agreement with the FEM prediction. Some bouncing of the acoustic wave inside the waveguide is observed, which we attribute to imperfect mode excitation at the source and to acoustic wave walk-off caused by the anisotropic nature of lithium niobate.

\section{IDT pair S-parameter measurement}\label{smsec:idt_pair}

To characterize the propagation of acoustic wave in the device, we perform vector network analyzer (VNA) measurements on a pair of interdigital transducers (IDTs) arranged in a face-to-face configuration. Each IDT is connected to one port of the VNA through RF probes and coaxial cables. When an RF signal is applied to port 1, the first IDT converts the electrical signal into a propagating acoustic wave, which travels along the acoustic waveguide and is subsequently converted back into an electrical signal by the second IDT connected to port 2. The overall RF–acoustic–RF transduction process is captured by the complex transmission coefficient $S_{21}$ measured by the VNA.

The measured $S_{21}$ contains both amplitude and phase information associated with acoustic wave generation, propagation, and detection. When $S_{21}$ is transformed into the time domain via an inverse Fourier transform, the resulting impulse response can be interpreted as an acoustic “echo” signal, in which distinct peaks correspond to acoustic waves arriving at the receiving IDT after different round-trip times and, consequently, different propagation delays. 

By sweeping the separation between the IDT pairs, the temporal positions and pulse energies of the first arrival and subsequent echoes reveal key acoustic properties, including the acoustic group velocity and propagation loss. The detailed measurement procedure is as follows. First, we identify the frequency window corresponding to the excitation of the Rayleigh mode by the IDTs. The measured $S_{21}$ spectrum is then band pass filtered at this frequency range, and an inverse Fourier transform is applied to obtain the time-domain echo signal. From this signal, we extract the pulse energies of the first arrival and the subsequent echo, and calculate the energy loss between these two pulses.

\begin{figure}[htbp]
  \centering
  \includegraphics[width=\linewidth]{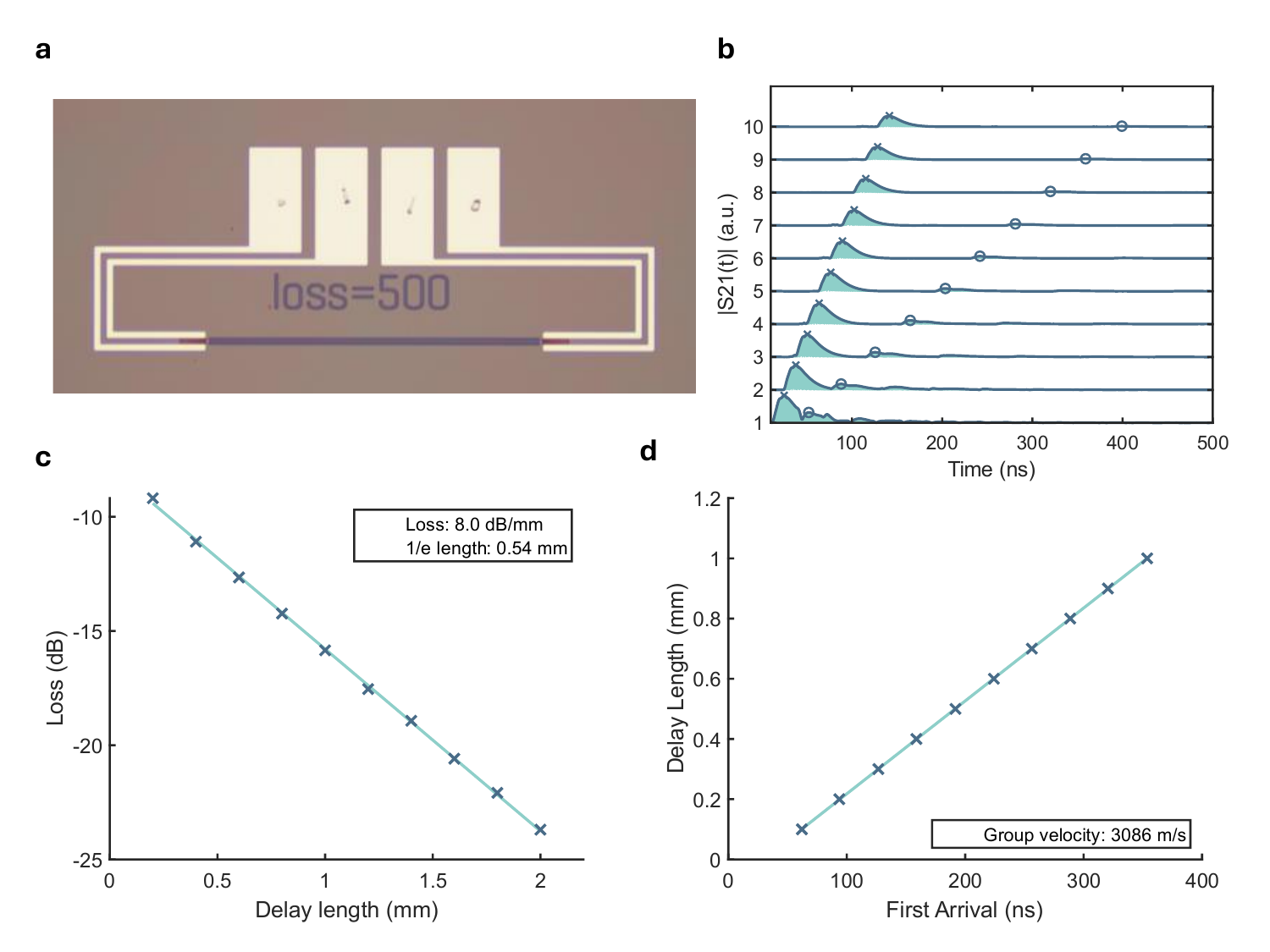}
  \caption{\textbf{Propagation loss and group velocity of Rayleigh wave.} (a) An IDT pair used in this measurement. The separation between the IDTs is $500\ \mathrm{\mu m}$, and the scratches at the top are caused by RF probes. (b) Time-domain $|S_{21}(t)|$ of IDT pairs with separations ranging from $100\ \mathrm{\mu m}$ to $1000\ \mathrm{\mu m}$ in steps of $100\ \mathrm{\mu m}$. The first arrival pulse and its subsequent echo are marked with crosses and circles, respectively. (c) Measured energy loss between the first arrival and its echo as a function of the delay length between them; the extracted propagation loss is $8.0\ \mathrm{dB/mm}$. (d) Measured IDT pair separation versus first-arrival time; the extracted group velocity is $3086\ \rm{m/s}$.}\label{smfig:idt_pair_loss}
\end{figure}

By varying the separation between the IDT pairs, the additional energy loss arises solely from the increased acoustic propagation distance between the first arrival and the echo. Therefore, by fitting the extracted energy loss as an exponential function of the corresponding propagation delay length, we obtain the propagation loss of the Rayleigh wave. In addition, the Rayleigh-wave group velocity is extracted by linearly fitting the IDT separation as a function of the first-arrival time. Based on this experimental scheme, we measured the propagation loss of the Rayleigh wave to be $8.0\ \mathrm{dB/mm}$ and its group velocity to be $3086\ \rm{m/s}$. Results are shown in Fig.~\ref{smfig:idt_pair_loss}. The measured  $8.0\ \mathrm{dB/mm}$ acoustic propagation loss is comparable to that of an unconfined slab Rayleigh SAW, indicating that the acoustic waveguide does not introduce significant additional propagation loss.

\begin{figure}[htbp]
  \centering
  \includegraphics[width=\linewidth]{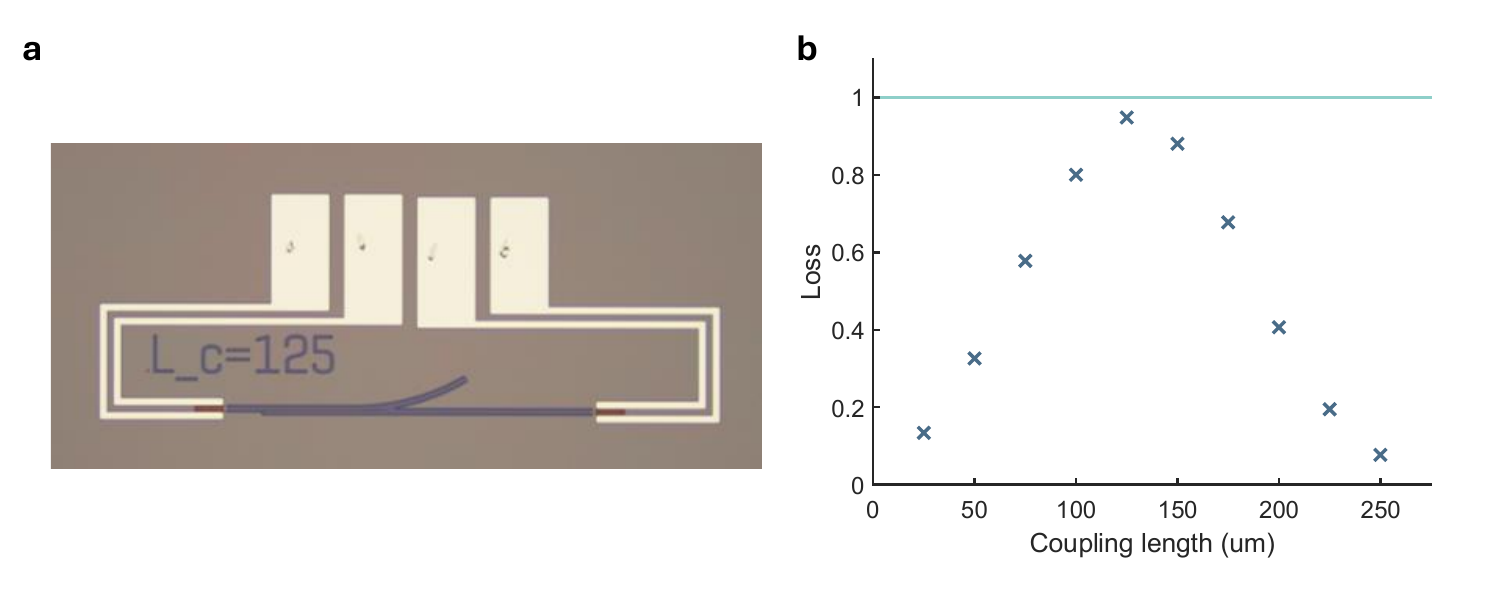}
  \caption{\textbf{Evanescent coupling of Rayleigh wave.} (a) An IDT pair used in this measurement. The separation between the IDTs is $500\ \mathrm{\mu m}$, the coupling length between the waveguides is $125\ \mathrm{\mu m}$, and the scratches at the top are caused by RF probes. (b) Measured power decay between the first arrival pulse and its subsequent echo as a function of waveguide coupling length. The coupling length ranges from $25\ \mathrm{\mu m}$ to $250\ \mathrm{\mu m}$ in steps of $25\ \mathrm{\mu m}$, and the total waveguide length between IDT pair is $500\ \mathrm{\mu m}$. The power decay is normalized to that from an IDT pair directly connected by a $500\ \mathrm{\mu m}$-long waveguide, and the corresponding unity power transmission is indicated by a horizontal line.}\label{smfig:idt_pair_couple}
\end{figure}

Similarly, the evanescent coupling ratio between two acoustic waveguides can be measured by keeping the IDT pair separation fixed while sweeping the coupling length. This approach eliminates the influence of propagation loss, allowing direct measurement of the energy transfer ratio between the waveguides. The results are shown in Fig.~\ref{smfig:idt_pair_couple}, exhibiting a good $\sin^2$-like behavior. The measured coupling length required to achieve near-unity power transfer is approximately $125\ \rm{\mu m}$, in good agreement with our simulations.

\section{Frequency-dependent analysis of rAOBS efficiency and saturation}\label{smsec:all_freq_measurement}

To further examine the origin of the saturation behavior in the measured rAOBS efficiency, we performed RF-power-dependent measurements at multiple acoustic frequencies across the IDT bandwidth. For each acoustic frequency and RF power, the optical resonance spectrum was measured and fitted to extract the loaded quality factor and the cavity round-trip loss. In parallel, the free-space AOBS efficiency was directly measured by collecting the upward-scattered optical beam in air.

\begin{figure}[htbp]
  \centering
  \includegraphics[width=\linewidth]{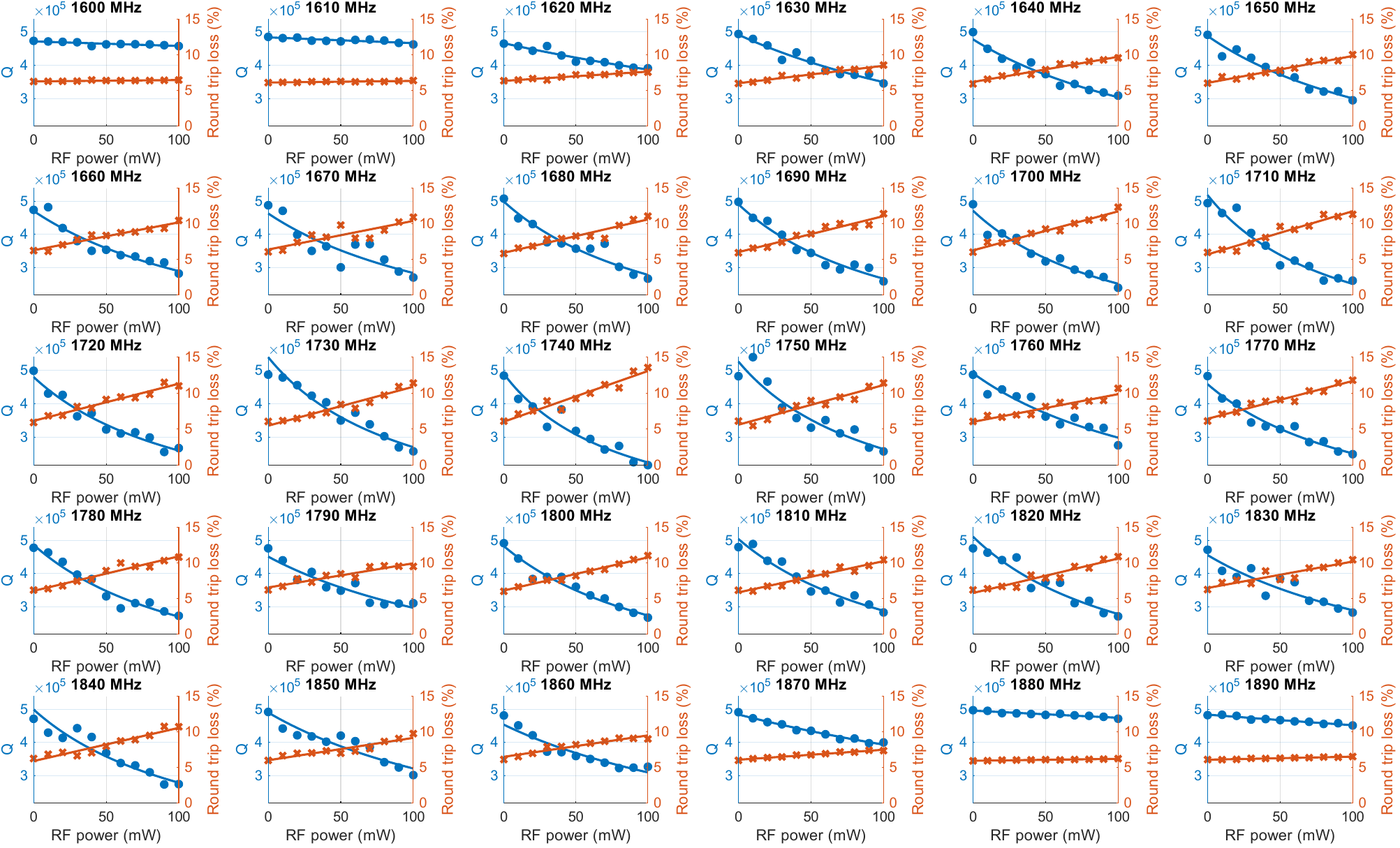}
  \caption{\textbf{RF-power-dependent loaded quality factor and round-trip loss at different acoustic frequencies.} Optical resonance spectra were measured at different RF powers for multiple acoustic frequencies across the IDT bandwidth. Each panel corresponds to one acoustic frequency. Circles and crosses represent the loaded quality factor and cavity round-trip loss extracted from the fitted resonance spectra, respectively. Solid lines are fits to the extracted data.}\label{smfig:q_loss_vs_freq}
\end{figure}

Fig.~\ref{smfig:q_loss_vs_freq} shows fitted loaded quality factor and cavity round-trip loss. The extracted cavity round-trip loss increases approximately linearly with RF power at all measured acoustic frequencies. This indicates that, within the measured RF power range, the RF-to-acoustic conversion of the IDT does not reach a saturation regime. Small fluctuations in the extracted values are mainly attributed to variations in the fiber-to-chip coupling background, which can introduce uncertainty in the resonance fitting. Nevertheless, the overall linear dependence of the extracted round-trip loss on RF power is consistently observed across the measured acoustic frequencies, including frequencies near the edge of the IDT bandwidth.

\begin{figure}[htbp]
  \centering
  \includegraphics[width=\linewidth]{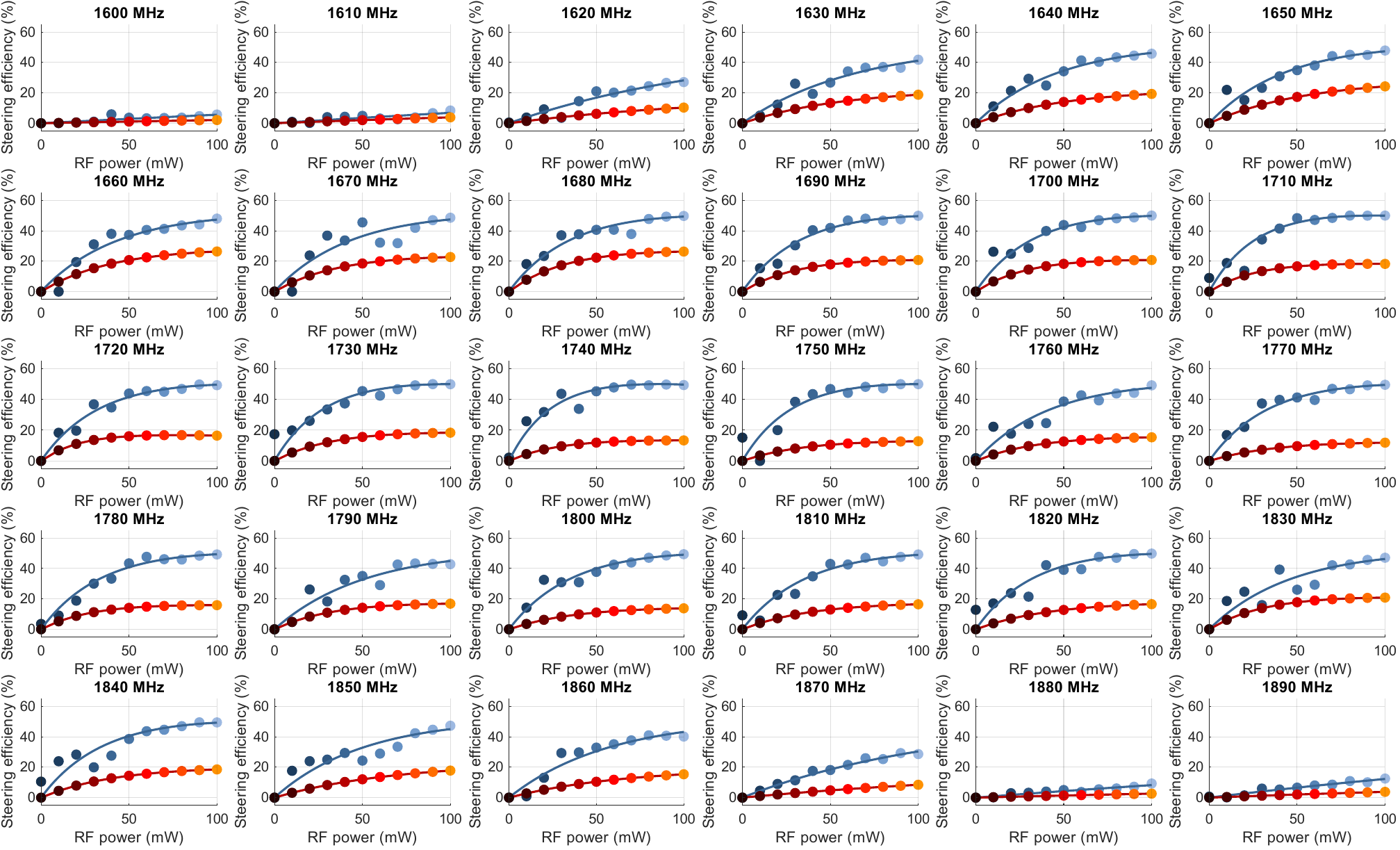}
  \caption{\textbf{RF-power-dependent rAOBS efficiency at different acoustic frequencies.} Each panel corresponds to one acoustic frequency. The blue curves show the total rAOBS efficiency inferred from resonance-spectrum fitting, including both free-space steering and substrate steering. The red curves show directly measured free-space steering efficiency.}\label{smfig:theory_meas_vs_freq}
\end{figure}

We then compare two quantities for the AOBS efficiency (Fig.~\ref{smfig:theory_meas_vs_freq}). The first is the total rAOBS efficiency inferred from the resonance-spectrum fitting, which accounts for the total acousto-optic scattering loss induced in the cavity, including both free-space steering and substrate scattering. The second is the directly measured free-space steering efficiency, which only includes the upward-scattered optical power collected in air. Although these two quantities have different absolute values due to the downward scattering component, both exhibit saturation behavior as the RF power increases.

This observation shows that the saturation of the rAOBS efficiency does not originate from saturation of the IDT or acoustic generation. Instead, as the acoustically induced scattering loss increases with RF power, the optical resonator transitions from the critically coupled regime toward the under-coupled regime. In this regime, further increasing the acousto-optic scattering loss no longer leads to a proportional increase in the externally measured steering efficiency, resulting in the observed saturation.

\begin{figure}[htbp]
  \centering
  \includegraphics[width=.6\linewidth]{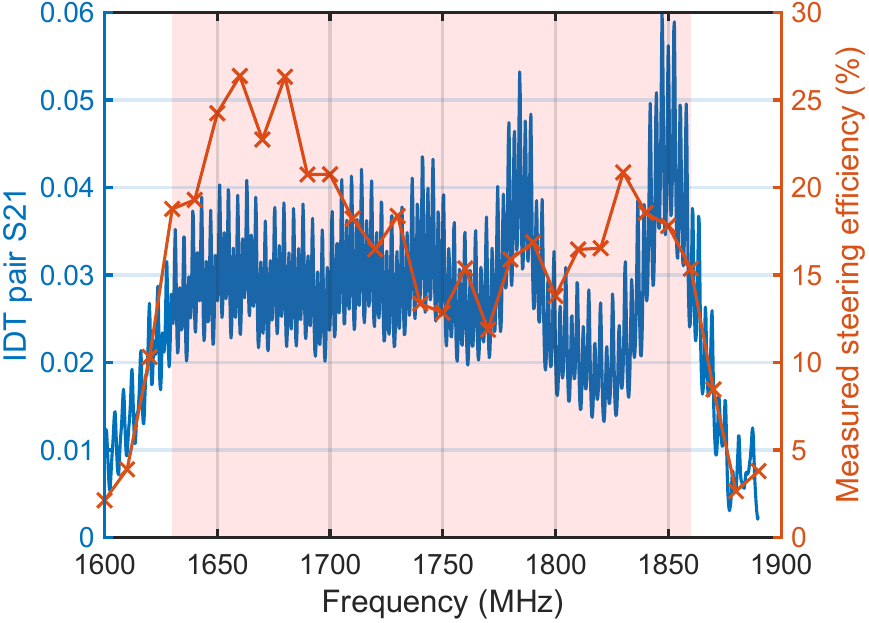}
  \caption{\textbf{Comparison between IDT pair $S_{21}$ response and measured steering efficiency.} The blue curve shows the RF transmission spectrum $S_{21}$ measured from a pair of identical IDTs. The orange curve shows the directly measured free-space rAOBS steering efficiency at 100 mW RF driving power, as a function of acoustic frequency. The shaded region indicates the acoustic-frequency range corresponding to the demonstrated field of view.}\label{smfig:eff_vs_idt}
\end{figure}

To identify the factor limiting the demonstrated field of view, we compared the free-space rAOBS steering efficiency at 100 mW RF driving power, with the RF transmission response of the acoustic transducers. The results are shown in Fig.~\ref{smfig:eff_vs_idt}. The RF transmission spectrum $S_{21}$ was measured between two identical IDTs, corresponding to an RF–acoustic–RF transduction process. Therefore, the $S_{21}$ response provides an estimate of the acoustic generation and coupling bandwidth of the IDTs. The oscillatory features observed in the $S_{21}$ spectrum originate from acoustic interference between the two IDTs.

The measured free-space steering efficiency follows the same overall frequency bandwidth as the IDT $S_{21}$ response. Efficient beam steering is observed only within the frequency range where the IDTs efficiently generate and couple acoustic waves into the resonator. This agreement indicates that the demonstrated $18^\circ$ field of view is limited by the available acoustic bandwidth of the IDTs, rather than by degradation of beam quality or optical aberration at the edge of the steering range.

\section{Experimental setup for electro-optic resonance locking}\label{smsec:locking}

\begin{figure}[htbp]
  \centering
  \includegraphics[width=.8\linewidth]{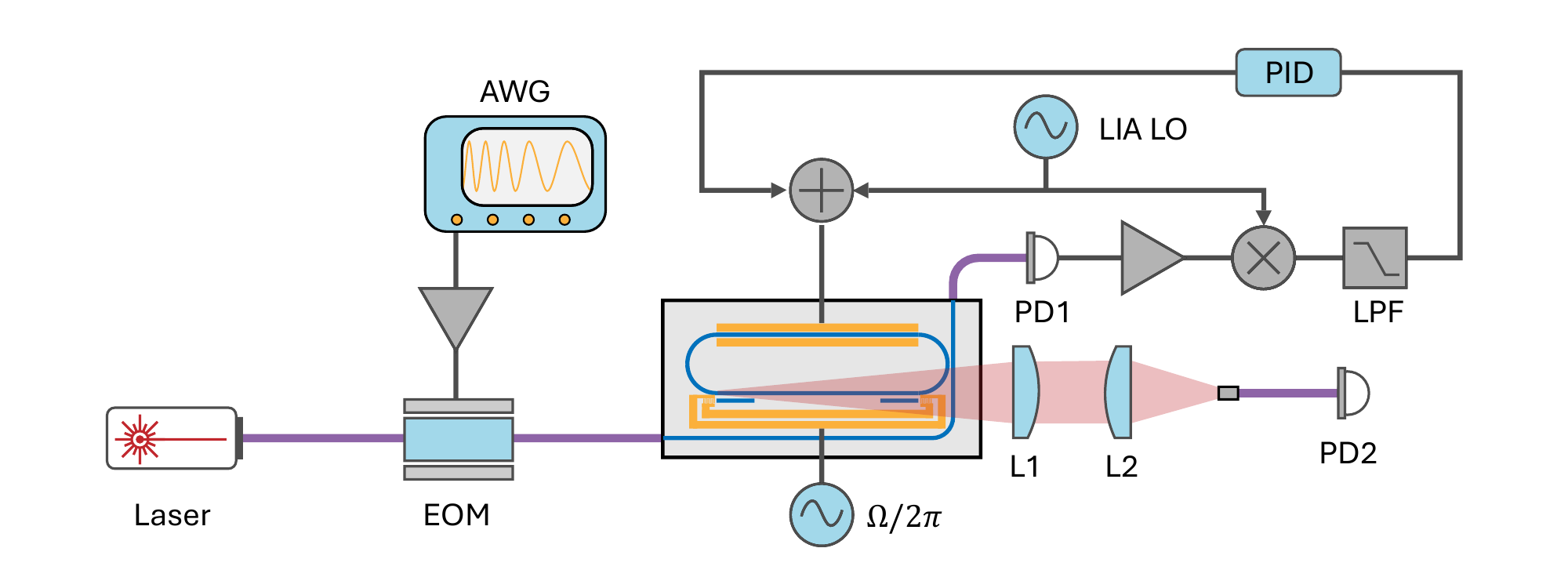}
  \caption{\textbf{Experimental setup for electro-optic resonance locking: direct PD measurement} PD1: output photodetector; LPF: low-pass filter; LIA LO: lock-in amplifier's local oscillator; PD2: free-space photodetector; L1: cylindrical plano-convex lens; L2: plano-convex lens}\label{smfig:locking_direct_pd}
\end{figure}

\begin{figure}[htbp]
  \centering
  \includegraphics[width=.8\linewidth]{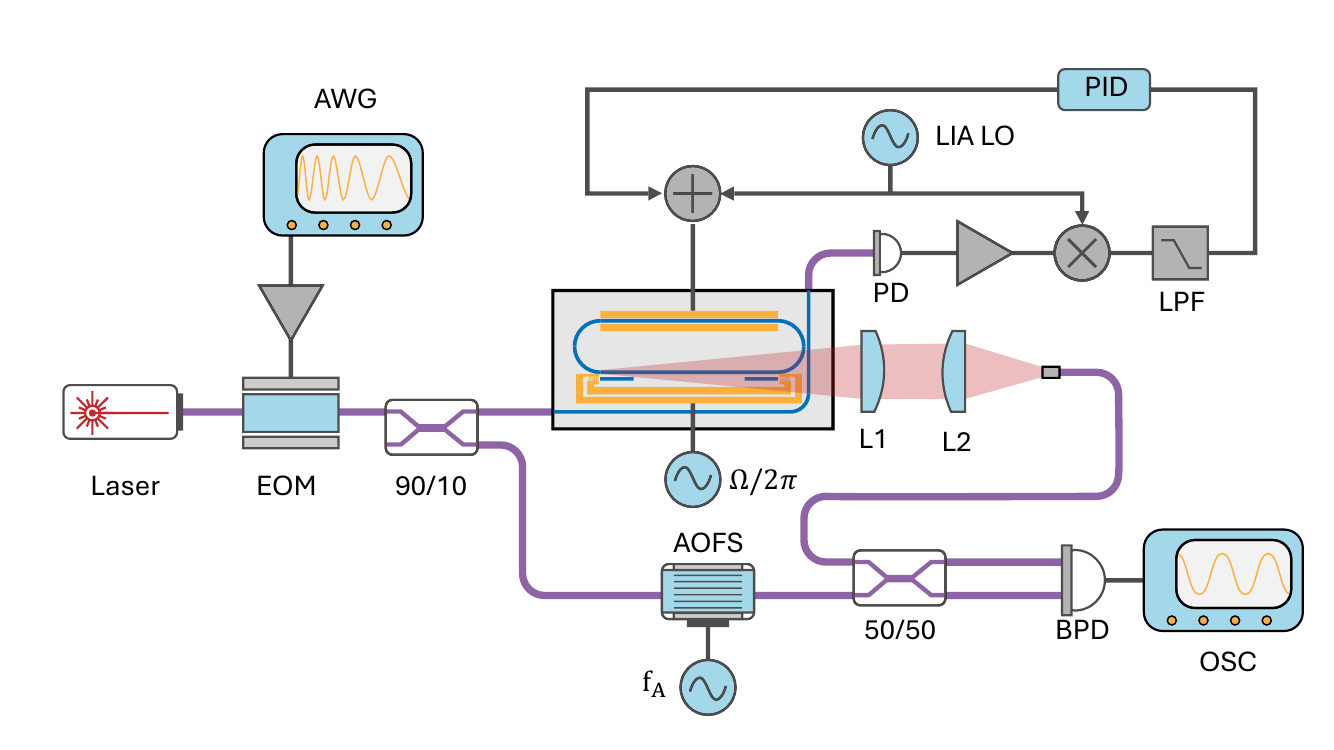}
  \caption{\textbf{Experimental setup for electro-optic resonance locking: heterodyne measurement} PD: output photodetector; LPF: low-pass filter; LIA LO: lock-in amplifier's local oscillator; BPD: balanced photodetector; L1: cylindrical plano-convex lens; L2: plano-convex lens}\label{smfig:locking_heterodyne}
\end{figure}

Figure~\ref{smfig:locking_direct_pd} shows the experimental setup for direct free-space beam power measurement. Fig.~\ref{smfig:locking_heterodyne} shows the setup used for heterodyne detection.

\section{Lock-in detection of resonance detuning}\label{smsec:slope}

\begin{figure}[htbp]
  \centering
  \includegraphics[width=\linewidth]{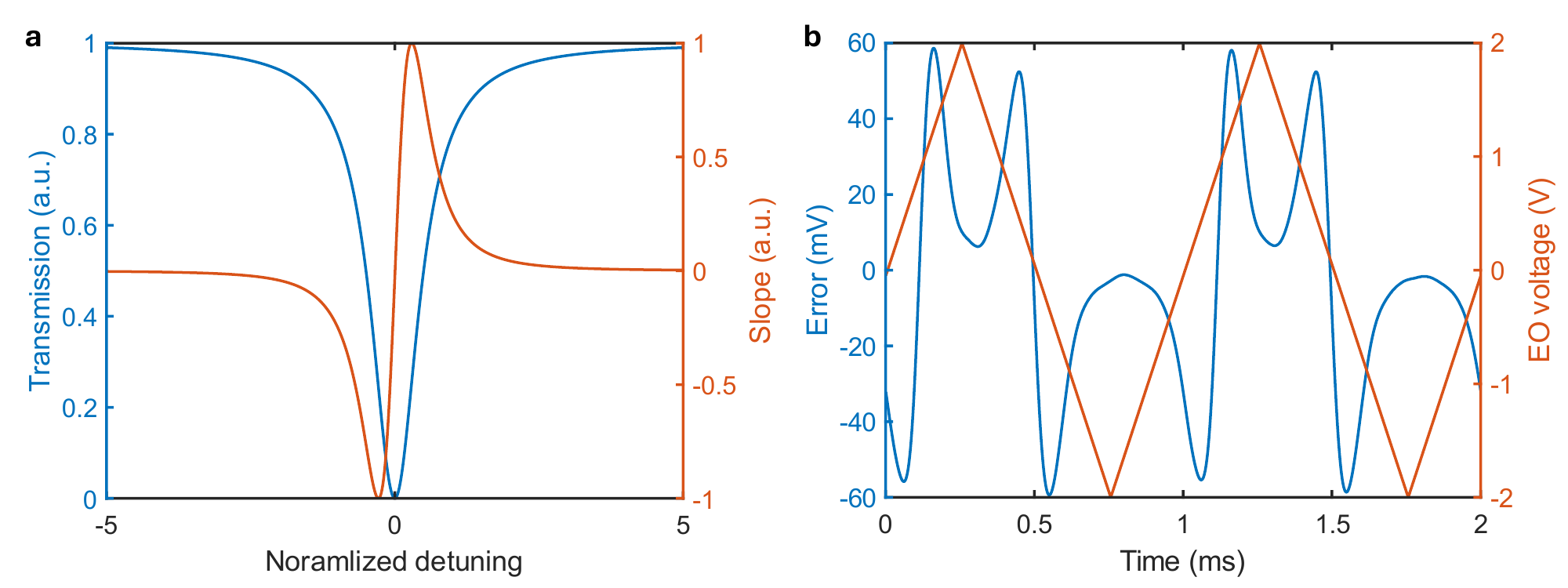}
  \caption{\textbf{Slope detection} (a) A Lorentzian resonance spectrum with unit linewidth is plotted in blue, and its slope is plotted in orange. The slope and detuning always have the same sign. (b) Measured error signal detected by the PID controller when a sawtooth EO voltage is applied to the optical cavity.}\label{smfig:slope}
\end{figure}

When the resonator is electro-optically perturbed by a weak RF signal, its resonance spectrum dithers, producing an optical transmission signal at the same frequency, which is detected by the output PD. The amplitude of the RF signal measured at the PD is proportional to the slope of the transmission spectrum at the detuning where it is perturbed, and the slope always has the same sign as the detuning, as illustrated in Fig.~\ref{smfig:slope}a. This allows the complex amplitude of the RF signal detected by the PD to be projected onto an axis in the complex plane by doing lock-in measurement, which can then serve as the input to the PID controller.

Fig.~\ref{smfig:slope}b shows a measured error signal detected by the PID controller when a sawtooth voltage is applied to the EO electrode to tune its resonance across the laser frequency. A clear pattern resembling the slope in Fig.~\ref{smfig:slope}a is observed, validating the analysis above.

\section{Resonance locking analysis and procedures}\label{smsec:pid}

In the FMCW LiDAR experiment, the optical frequency is chirped by $5~\mathrm{GHz}$ over each $0.5~\mathrm{ms}$ up- or down-chirp. Although this corresponds to a large optical-frequency sweep rate, the resonance-locking loop does not directly respond to the optical-frequency ramp in units of $\mathrm{GHz/ms}$. Instead, the loop acts on the voltage applied to the resonator EOM, which follows a periodic sawtooth waveform with a fundamental frequency of $1~\mathrm{kHz}$, together with residual detuning errors and noise. Therefore, the relevant feedback bandwidth is determined by the temporal frequency components of the EOM tuning waveform, rather than by the absolute optical-frequency sweep rate.

The stated $10~\mathrm{kHz}$ feedback bandwidth refers to the low-pass filtering bandwidth of the mixer IF signal used in the lock-in detection before the PID controller. Thus, the PID controller processes error signals below $10~\mathrm{kHz}$. This bandwidth is sufficient to include the $1~\mathrm{kHz}$ chirp waveform and its relevant higher-order harmonics for the present FMCW operation. Experimentally, we verified that the resonance can be tracked using feedback alone under the chirp condition used in this work.

In addition to feedback-only operation, a sawtooth feedforward signal can be applied simultaneously to improve the robustness and stability of the resonance locking. Since the required chirp waveform is deterministic and periodic, feedforward compensation can provide most of the large-amplitude EOM tuning needed to follow the chirped laser, while the feedback loop corrects the residual detuning error. This reduces the voltage swing and open-loop gain required from the feedback path, thereby improving the stability margin and reducing sensitivity to high-frequency poles and phase delay in the feedback loop.

\begin{table}[htbp]
\caption{\textbf{Key parameters of the resonance-locking and FMCW chirp-control system.} The table summarizes the relevant chirp parameters, electro-optic tuning coefficient, and control parameters used for resonance tracking during resonance locking operation.}
\label{smtab:feedback}
\centering
\begin{tabular}{ll}
\hline
Item & Value \\
\hline
Chirp period & $1~\mathrm{ms}$ \\
Up/down chirp duration & $0.5~\mathrm{ms}$ \\
Chirp excursion & $5~\mathrm{GHz}$ \\
Resonator EOM tuning coefficient & $0.3~\mathrm{GHz/V}$ \\
Resonator EOM-to-PID voltage gain (plant gain) & $0.15$ \\
Required resonator EOM voltage swing & $5/0.3 \approx 17~\mathrm{V}$ \\
Target residual detuning & 0$.1~\mathrm{GHz}$ \\
Target residual EOM voltage & $0.1/0.3 \approx 0.3~\mathrm{V}$ \\
\hline
\end{tabular}
\end{table}

The relevant parameters for the resonance-locking and FMCW chirp-control system are summarized in Table~\ref{smtab:feedback}. The resonator EOM tuning coefficient is approximately $0.3~\mathrm{GHz/V}$. Therefore, tracking a $5~\mathrm{GHz}$ chirp requires an EOM voltage swing of approximately
\begin{equation}
    V_\mathrm{chirp} \approx \frac{5~\mathrm{GHz}}{0.3~\mathrm{GHz/V}} \approx 17~\mathrm{V}.
\end{equation}

For a target residual detuning of $0.1~\mathrm{GHz}$, the corresponding residual EOM voltage is
\begin{equation}
    V_\mathrm{res} \approx \frac{0.1~\mathrm{GHz}}{0.3~\mathrm{GHz/V}} \approx 0.3~\mathrm{V}.
\end{equation}

Thus, the required suppression ratio between the full chirp voltage and the residual error voltage is approximately
\begin{equation}
    G\sqrt{K_p^2+\left(\frac{K_i}{2\pi f}\right)^2}\gtrsim 57
\end{equation}

The following are the exact step-by-step operations:

\begin{enumerate}
\item Turn on the AWG to generate a single-frequency tone. The frequency should correspond to the center frequency of the chirp waveform that will be used in FMCW LiDAR.
\item Scan the wavelength of the laser to identify the optimal resonance dips. Each dip family contains three dips, corresponding to the input laser wavelength and its upper and lower sidebands generated by the EOM.
\item Park the laser wavelength at one of the sidebands.
\item Reconfigure the AWG to generate a sawtooth-frequency-chirp waveform to chirp the input laser, thereby generating two chirped sidebands.
\item (Optional) To facilitate the locking between the resonator and the laser sideband, an additional sawtooth waveform, in-phase with the sawtooth-frequency-chirp waveform driving the EOM, can be generated by the same AWG. This waveform can be amplified and applied to the EO electrodes of the resonator to compensate for the predictable laser frequency sweep. This method reduces the required feedback gain, thereby lowering the risk of feedback instability and oscillation.
\item Turn on the proportional component of the PID controller, gradually increasing it from zero until the resonator transmission becomes unstable. Once instability occurs, gradually decrease the proportional gain until the transmission stabilizes. Note that “stable” here indicates the absence of feedback oscillation, not necessarily that the resonator tracks the chirped sideband.
\item Turn on the integral component of the PID controller and gradually increase it from zero. The resonator will then track the chirped sideband, and the voltage applied to the EO electrodes will exhibit a sawtooth waveform, which can be monitored using an oscilloscope.
\end{enumerate}

\section{Optical setup for focal plane measurement}\label{smsec:k_space}

\begin{figure}[htbp]
  \centering
  \includegraphics[width=.6\linewidth]{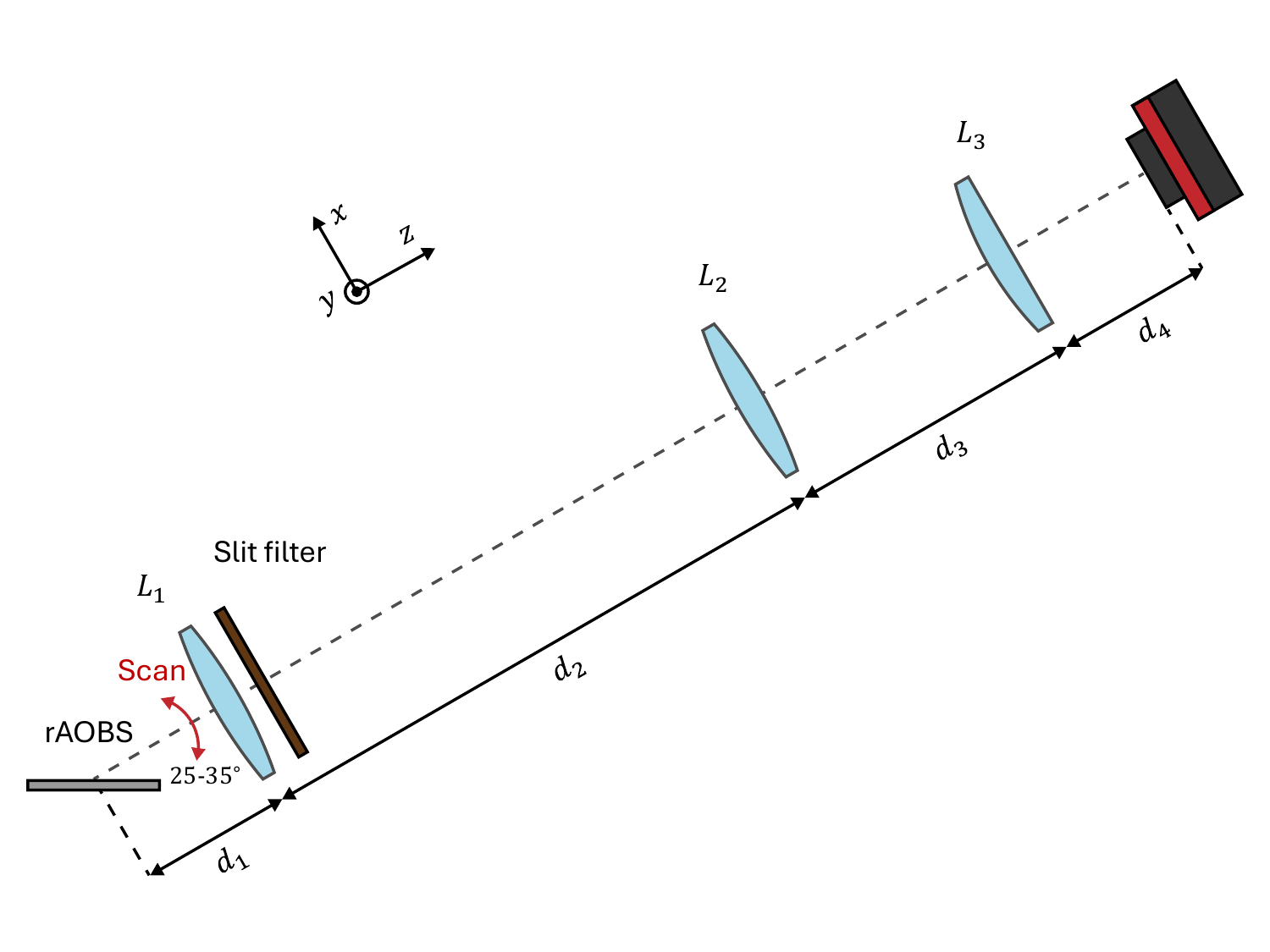}
  \caption{\textbf{Optical setup for focal-plane measurement.} $L_1$ is a bi-convex lens with a focal length of $50\ \mathrm{mm}$. $L_2$ is a bi-convex lens with a focal length of $150\ \mathrm{mm}$, and $L_3$ is a cylindrical plano-convex lens with its cylinder axis aligned along the $y$ axis and a focal length of $50\ \mathrm{mm}$. A slit filter is placed close to $L_1$, with its slit oriented parallel to the $x$ axis. Distances $d_1$ to $d_4$ are $50\ \mathrm{mm}$, $200\ \mathrm{mm}$, $100\ \mathrm{mm}$, and $50\ \mathrm{mm}$, respectively. The beam-steering direction of the rAOBS chip is indicated by the red arrow. The angle between the optical axis of the imaging system and the horizontal plane can be adjusted between $25^\circ$ and $35^\circ$ to minimize off-axis imaging aberration.}\label{smfig:k_space_setup}
\end{figure}

The optical setup for the focal-plane measurement is shown in Fig.~\ref{smfig:k_space_setup}. The beam emitted into free space is divergent in the direction perpendicular to the scan direction due to the finite width of the photonic waveguide. As a result, the beam is collimated along the scan direction ($x$ axis) but diverges along the $y$ axis. Therefore, to convert the steered beam into a scanning spot for focal-plane measurement, the optical system must treat the $x$ and $y$ axes differently.

Along the $x$ axis, which corresponds to the beam-steering direction, lens $L_1$ is used to map the collimated output beam into its angular (momentum) space at the back focal plane of $L_1$. Lenses $L_2$ and $L_3$ then image and demagnify this angular space distribution by a factor of $1/3$ so that it fits within the sensing area of the infrared camera.

Along the $y$ axis, the objective is to first collimate the divergent beam and then focus it onto the camera’s sensing area. This is achieved using only lenses $L_1$ and $L_2$. Lens $L_3$ is a cylindrical lens with its cylinder axis aligned along the $y$ axis, and therefore it does not affect the beam propagation in the $y$ direction.

A slit filter is placed after $L_1$, with its slit oriented parallel to the $x$ axis. The filter blocks beams that are excessively divergent along the $y$ axis, which would otherwise degrade the beam profile in the focal plane.

\section{Effect of optical-power ripple on FMCW ranging accuracy}\label{smsec:pid_effect}

In the FMCW LiDAR experiment, the chirped optical signal is generated using an EOM driven by an RF waveform from an AWG. In practice, the finite frequency response of the EOM and spurious signals from the AWG can introduce optical-power ripples during the chirp. These effects may modulate the amplitude of the transmitted and received optical signals and could, in principle, affect the ranging accuracy or resolution.

To evaluate this effect, we used the experimentally measured optical power envelope of the rAOBS-steered chirped beam during resonance locking in Fig.~\ref{fig3}c as the input to a numerical FMCW LiDAR simulation. This measured envelope includes the combined influence of the EOM frequency-response ripple, AWG waveform imperfections, and resonance-locking-related power variation. The measured envelope was used to modulate the optical power of the simulated local oscillator and received signal. The simulated target distance was set to 2 m, and the frequency separation between the local oscillator and received signal was set to 1.7 GHz, corresponding to the acoustic frequency used in the rAOBS measurement.

\begin{figure}[htbp]
  \centering
  \includegraphics[width=.8\linewidth]{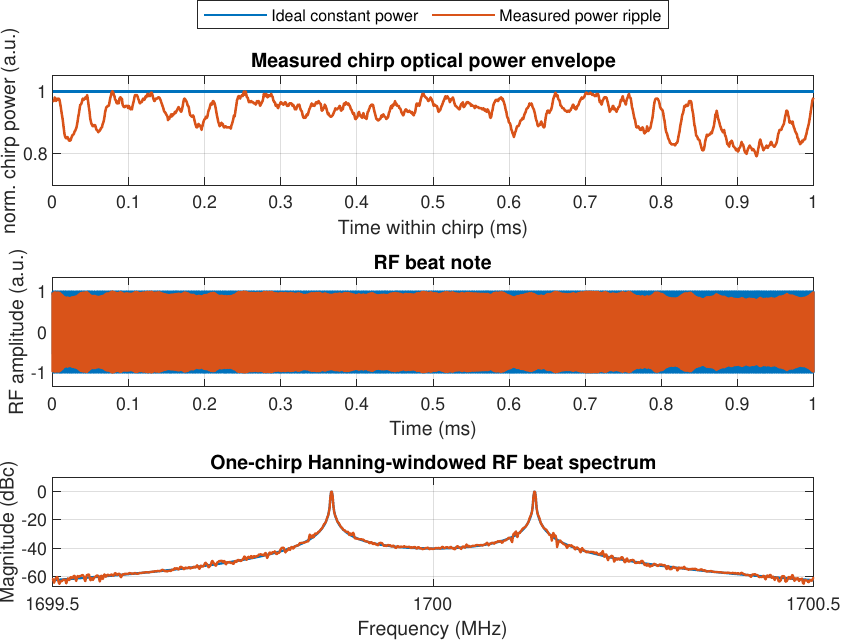}
  \caption{\textbf{Effect of optical-power ripple on FMCW ranging.} Simulated FMCW LiDAR chirp power, RF beat note, and RF beat spectrum comparing ideal constant optical power with measured chirp power ripple. The measured power envelope is extracted from the PID-on free-space photodiode signal and applied to both the local oscillator and delayed return fields, while the chirp phase remains ideal. The spectrum is shown around the 1.7 GHz LO offset for a 2 m target range.
}\label{smfig:pid_effect}
\end{figure}

Fig.~\ref{smfig:pid_effect} compares the simulated RF beat note and its spectrum for two cases: an ideal chirped signal with constant optical power and a chirped signal with the experimentally measured power ripple. In the time domain, the measured optical-power ripple introduces a clear envelope modulation of the RF beat note. This modulation is expected because the heterodyne beat amplitude is proportional to the product of the local-oscillator and received optical field amplitudes. However, in the frequency domain, the beat-note spectra near the ranging frequency are nearly identical for the two cases. In particular, the beat-note peak position and linewidth show no noticeable change compared with the ideal constant-power case.

This weak impact on ranging accuracy can be understood from the different frequency scales involved. The optical-power ripple mainly varies on the chirp-period time scale, approximately 1 ms in the present experiment, and therefore acts as a low-frequency amplitude modulation of the RF beat note. In contrast, the ranging information is encoded in the instantaneous FMCW beat frequency, which appears as a high-frequency spectral peak around the acoustic-frequency-shifted beat note. Since the slow optical-power modulation does not appreciably change the instantaneous beat frequency, it primarily modifies the beat-note envelope and weakly affects the spectral background, while leaving the ranging peak position and linewidth nearly unchanged.

\section{Analysis of ranging performance and target detectability
}\label{smsec:lidar_discussion}

The ranging resolution in the current FMCW LiDAR experiment is limited by the Fourier-transform frequency resolution of the measured beat signal~\cite{kunita2007range, ko2008range}. In our measurement, both the up-chirp and down-chirp durations are 0.5 ms, corresponding to a Fourier frequency bin size of 2 kHz. Combined with the 5 GHz chirp excursion, this gives a minimum resolvable distance of approximately 3 cm. This estimate is consistent with the experimentally measured FWHM of the beat-note peak in Fig.~\ref{fig4}e. Further improvement of this resolution can be achieved by increasing the frequency chirp excursion.

The minimum resolvable target size by a single laser beam is primarily determined by the optical beam divergence. At a target distance $d$, the beam spot size on the target is approximately given by the product of the beam divergence angle and $d$, which limits the minimum target dimension that can be resolved, although advanced Gaussian decomposition and reconstruction algorithms can improve this limit to some extent. For our rAOBS devices, the experimentally measured beam divergence angles are $0.180^\circ$ and $0.030^\circ$ parallel and perpendicular to the steering direction, respectively, as shown in Fig.~\ref{fig2}b. Assuming a target distance of 10 m, these divergence angles correspond to beam dimensions of approximately 31 cm along the steering direction and 5.2 cm in the orthogonal direction.

\begin{figure}[htbp]
\centering
   \includegraphics[width=.6\linewidth]{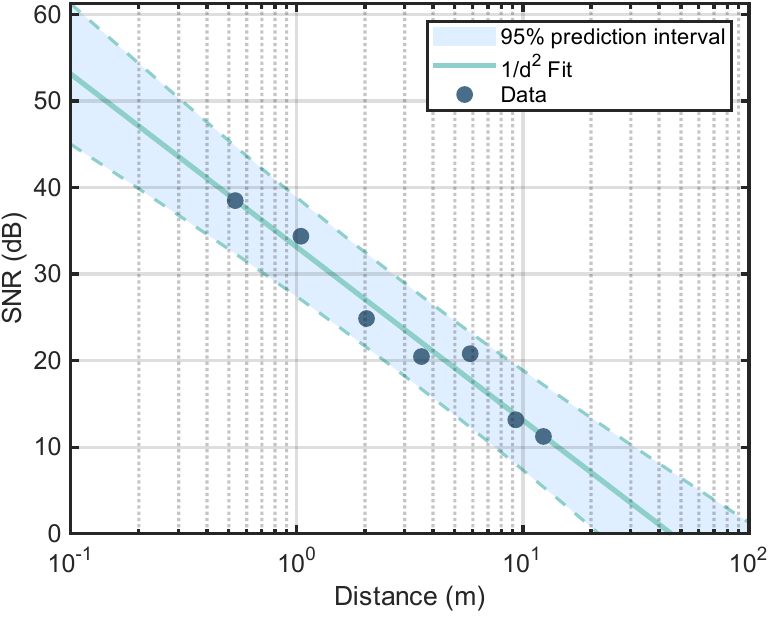}
\caption{\textbf{Experimental estimation of the maximum ranging distance.} Measured FMCW beat-note SNR as a function of target distance, obtained with a chirped steered optical power of approximately $1~\mathrm{mW}$. The solid line shows the $1/d^2$ fitted model used to estimate the maximum ranging distance, and the shaded region indicates the corresponding $95\%$ confidence interval.}\label{smfig:lidar_range}
\end{figure}

To characterize the ranging performance of the rAOBS-based FMCW LiDAR system, we measured the heterodyne beat-note signal from a target placed at different distances up to 12 m, using a steered beam power of approximately 1 mW. The target was made using retroreflective tape to provide a stable and reproducible optical return. At each target distance, the FMCW beat spectrum was recorded and the signal-to-noise ratio (SNR) of the ranging peak was extracted. The measured SNR decreases with increasing target distance. The experimental data were fitted using a $1/d^2$ model for a diffuse target and the corresponding $95\%$ prediction interval is shown together with the measured data in Fig.~\ref{smfig:lidar_range}. From this plot, we extract a ranging distance of 40 m at 3dB SNR. The ranging distance can be further increased by further improving system SNR, including increasing receiver aperture size, using operating the photodetector at the shot-noise limit, and increasing the output beam power. For exmaple, if output power can be increased to 10 mW, this ranging distance can be increased by $\sqrt{10}$  times to 126 m. 


Compared with our previous slab-AOBS device~\cite{li2023frequency}, although the beam steering efficiency has been significantly improved, the beam divergence has increased and becomes a limiting factor in the LiDAR system's SNR. The higher divergence is due to the use of waveguide-guided acoustic waves in the present device. Approaches to mitigate this issue so as to further improve the system SNR and ranging distance  are still being explored.

\end{document}